\documentclass[fleqn,usenatbib]{mnras}

\usepackage[T1]{fontenc}
\usepackage{ae,aecompl}

\usepackage{graphicx}	% Including figure files
\usepackage{amsmath}	% Advanced maths commands
\usepackage{amssymb}	% Extra maths symbols
\usepackage{multirow}

\newcommand{\hMsun}{h^{-1}\mathrm{M_\odot}}
\newcommand{\hMpc}{h^{-1}\mathrm{Mpc}}

\newcommand{\Mpc}{\mathrm{Mpc}}
\newcommand{\sqdeg}{\mathrm{deg}^2}
\newcommand{\persqdeg}{\mathrm{deg}^{-2}}
\newcommand{\red}{}

\title[DESI BGS Incompleteness]{Correcting for Fibre Assignment Incompleteness in the DESI Bright Galaxy Survey}

\author[A. M. J. Smith et al.]
{Alex Smith,$^{1,2}$\thanks{E-mail: \red{alexander.smith@cea.fr}}
Jian-hua He,$^{2}$
Shaun Cole,$^{2}$
Lee Stothert,$^{2,3}$
Peder Norberg,$^{2,3}$
\newauthor
Carlton Baugh,$^{2}$
Davide Bianchi,$^{4}$
Michael J. Wilson,$^{5}$
David Brooks,$^{6}$
\newauthor
Jaime E. Forero-Romero,$^{7}$
John Moustakas,$^{8}$
Will J. Percival,$^{4,9,10}$
Gregory Tarle,$^{11}$
\newauthor
and Risa H. Wechsler$^{12,13}$
\\
$^{1}$\red{IRFU, CEA, Universit\'e Paris-Saclay, F-91191 Gif-sur-Yvette, France} \\
$^{2}$Institute for Computational Cosmology, Dept. of Physics, Univ. of Durham, South Road, Durham DH1 3LE, UK \\
$^{3}$Centre for Extragalactic Astronomy, Dept. of Physics, Univ. of Durham, South Road, Durham DH1 3LE, UK \\
$^{4}$Institute of Cosmology \& Gravitation, University of Portsmouth, Dennis Sciama Building, Portsmouth, PO1 3FX, UK \\
$^{5}$Lawrence Berkeley National Laboratory, 1 Cyclotron Road, Berkeley, CA 94720, USA \\
$^{6}$Department of Physics \& Astronomy, University College London, Gower Street, London, WC1E 6BT, UK\\
$^{7}$Departamento de F\'isica, Universidad de los Andes, Cra. 1 No. 18A-10, Edificio Ip, CP 111711, Bogot\'a, Colombia \\
$^{8}$Department of Physics and Astronomy, Siena College, 515 Loudon Road, Loudonville, NY 12211, USA \\
$^{9}$Department of Physics and Astronomy, University of Waterloo, 200 University Ave W, Waterloo, ON N2L 3G1, Canada \\
$^{10}$Perimeter Institute for Theoretical Physics, 31 Caroline St. North, Waterloo, ON N2L 2Y5, Canada \\
$^{11}$Physics Department, University of Michigan Ann Arbor, MI 48109, USA \\
$^{12}$Kavli Institute for Particle Astrophysics and Cosmology and SLAC National Accelerator Laboratory, Menlo Park, CA 94305, USA \\
$^{13}$Physics Department, Stanford University, Stanford, CA 93405, USA
}

\date{Accepted XXX. Received YYY; in original form ZZZ}

\pubyear{2018}

\begin{document}
\label{firstpage}
\pagerange{\pageref{firstpage}--\pageref{lastpage}}
\maketitle

\begin{abstract}
The Dark Energy Spectroscopic Instrument
(DESI) Bright Galaxy Survey (BGS) will be a survey of bright, low redshift galaxies, which is
planned to cover an area of $\sim 14,000$ sq deg in 3 passes.
Each pass will cover the survey area with $\sim 2000$ pointings, each of area $\sim 8$ sq deg.
The BGS is currently proposed to consist of
a bright high priority sample to an $r$-band magnitude limit $r \sim 19.5$, with a 
fainter low priority sample to $r \sim 20$.
The geometry of the DESI fibre positioners 
in the focal plane of the telescope affects the completeness of the survey, and 
has a non-trivial impact on clustering measurements.
Using a BGS mock catalogue, we show that completeness due to fibre assignment
primarily depends on the surface density of galaxies. Completeness is high (>95\%) 
in low density regions, but very low (<10\%) in the centre of massive clusters.
\red{We apply the pair inverse probability (PIP) weighting} correction to clustering measurements
from a BGS mock which has been through the fibre assignment algorithm. This method is only
unbiased if it is possible to observe every galaxy pair.
To facilitate this, 
we randomly promote a small fraction of the fainter sample
to be high priority, and dither the set of tile positions by a small angle.
We show that inverse pair weighting combined with angular upweighting
provides an unbiased correction to galaxy clustering measurements
for the complete 3 pass survey, and also after 1 pass, which is highly incomplete.

\end{abstract}

% to an $r$-band magnitude
%limit of $r=20$.
%The BGS will cover $\sim 14,000$ sq deg in 3 passes, with two priority tiers of
%galaxies, based on apparent magnitude. 


%%%%%%%%%%%%%%%%%%%%%%%%%%%%%%%%%%%%%%%%%%%%%%%%%%

%%%%%%%%%%%%%%%%% BODY OF PAPER %%%%%%%%%%%%%%%%%%

\section{Introduction}
The Dark Energy Spectroscopic Instrument (DESI)~\citep{DESI2016a, DESI2016b} 
will conduct a large spectroscopic survey with the primary science
aims of making precision measurements of the baryon acoustic oscillation (BAO) 
scale and the large scale redshift space distortion (RSD) of galaxy clustering. 
BAO will be used to measure the expansion history of the Universe and constrain 
dark energy \citep[e.g.][]{Seo2003}. RSD will be used to measure
the growth rate of structure in the Universe, and place constraints on theories of 
modified gravity \citep[e.g.][]{Guzzo2008}. These measurements are complementary,
as they can be used to break degeneracies between models of dark energy and RSD.
The instrument, which is nearing completion, % currently being built, 
will be installed on the 4-m Mayall Telescope at Kitt Peak, Arizona. 
DESI will consist of dark-time and bright-time programs. The dark-time survey
will measure spectra of 4 million luminous red galaxies (LRGs) ($0.4<z<1.0$), 17 million 
emission line galaxies (ELGs) ($0.6<z<1.6$), 1.7 million quasars ($z<2.1$) 
and 0.7 million high redshift quasars ($2.1<z<3.5$) to probe the
Ly-$\alpha$ forest.
The bright-time survey will consist of the bright galaxy survey
(BGS), a low redshift, flux limited survey
of $\sim$ 10 million galaxies with a median redshift $z_\mathrm{med} \sim 0.2$ 
(BGS paper, in prep), and a survey of Milky Way stars \citep{DESI2016a}.

The light from each target galaxy is collected by fibres located at the focal plane
of the telescope, and taken to one of 10 spectrographs, where the spectrum 
is measured and a redshift determined. However, it is not possible to place a 
fibre on every single potential target, and even if it is, a redshift measurement
can fail due to low surface brightness or weak spectral features. Other complications, such as 
observing conditions, also affect the redshift completeness in the final galaxy catalogue.\footnote{Exposures 
are scaled dynamically with conditions, with the aim of achieving
a consistent signal-to-noise ratio in the spectra.} To make
precise measurements of galaxy clustering in order to reach the primary science
aims of the survey, it is essential to correct for the effects of incompleteness.

A major systematic in galaxy clustering measurements is from the effect of fibre collisions,
which occur because fibres cannot be placed arbitrarily close together.
Since it is not possible to place a fibre on both objects in a close pair, 
that pair will be missing in the final catalogue,
biasing the pair counts, particularly at small scales, 
which can bias galaxy clustering measurements. If the
same patch of sky is observed enough times, the missing galaxies will eventually be observed,
removing the bias \citep[e.g. in GAMA][]{Robotham2010}, but typically it is infeasible to do this.

In the Sloan Digital Sky Survey (SDSS) \citep{Abazajian2009}, 
the fibre collisions can be characterised 
relatively straightforwardly, since fibres can be placed anywhere on a plate, so long as
they are not closer than the fibre collision scale of 55~arcsec (or 62~arcsec for BOSS).
A common method to recover
the redshift of missing galaxies is to simply assign them the same redshift as the nearest
targeted object on the sky \citep[e.g.][]{Zehavi2005,Zehavi2011}. 
However, this method produces unsatisfactory results
for the redshift-space correlation function (as shown in Section~\ref{sec:mitigation_comparison}).
An alternative method that works well for SDSS involves recovering the full correlation 
function from the regions covered by multiple overlapping tiles~\citep{Guo2012}. 
%This
%method assumes that the clustering in overlapping regions is representative, which is approximately
%true for the SDSS tiling, given the large fraction of overlapping regions, but not for the BGS.
In dense regions, SDSS is able to target all galaxies, or an unbiased subset, but
this is not true for the BGS.

Fibre collisions in DESI are more complicated, since the fibres are controlled by
robotic fibre positioners, which can move each fibre anywhere in a small patrol region
around a fixed set of centres, arranged in a grid. The fibre positioners can block neighbouring 
fibres from targeting certain objects, and objects will be missed if the number density of 
targets in an extended region is greater 
than the number density of fibres. These effects have a non-trivial impact on 
clustering estimates. The statistics to be measured from the survey can be modified to 
remove the affected scales \citep[e.g.][]{Burden2017,Pinol2017}, but in doing so,
information is lost. \citet{Bianchi2017} have proposed a method to correct clustering measurements 
by estimating, from many runs of the fibre assignment algorithm,
the probability that a pair of galaxies will be targeted, and have
shown that this method can provide an unbiased correction to the dark-time
ELG sample \citep{Bianchi2018}. The method has also been shown to be
successful when applied to data from the VIPERS survey \citep{Mohammad2018}.

Galaxies in the BGS have a variety of properties, and cover a wide range of 
galaxy bias. Many kinds of galaxy samples can be selected from the survey, such as volume
limited samples, stellar-mass selected samples and colour-selected samples.
Here, we quantify the incompleteness due to fibre assignment in the DESI
BGS, and assess correlation function correction techniques applied to 
samples from a BGS mock catalogue.
This paper is organised as follows: in Section~\ref{sec:fib_assign}, we describe the 
BGS survey strategy, DESI fibre assignment, and mock survey simulations. 
In Section~\ref{sec:completeness}, we quantify galaxy incompleteness in the BGS due
to fibre assignment. In Section~\ref{sec:corrections}, we assess correlation function 
correction methods on volume limited samples from the BGS mock.
Section~\ref{sec:conclusions} summarises our conclusions. Throughout, we
assume the WMAP-1 cosmology of the mock catalogue presented in Section~\ref{sec:surveysims}, 
with $\Omega_\mathrm{m}=0.25$, $\Omega_\Lambda=0.75$, $\sigma_8=0.9$,
$h=0.73$, and $n=1$ \citep{Spergel2003}.
While this cosmology has a higher $\sigma_8$ and lower $\Omega_\mathrm{m}$ than measurements
from Planck \citep{Planck2018}, we use simulations tuned to produce the correct
galaxy clustering, so we expect the dependence of our results on cosmology to be small.

\section{Fibre Assignment}
\label{sec:fib_assign}

\subsection{Survey Strategy}

\begin{figure} 
\centering
\includegraphics[width=9.3cm]{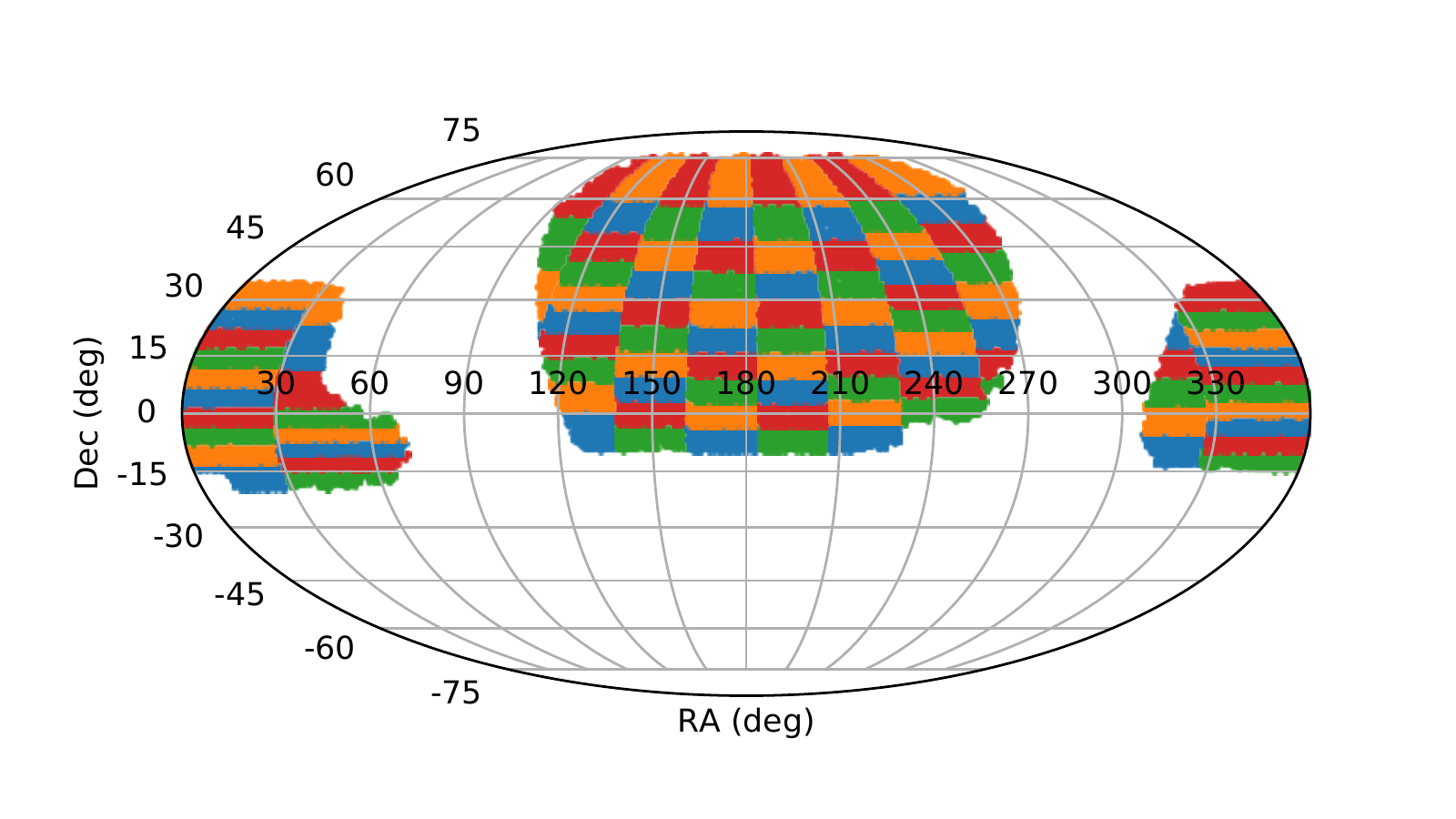}
\vspace{-0.7cm}
\caption{Footprint of the DESI BGS, which covers 14,800 square degrees. Colours
indicate the 100 jackknife regions.}
\label{fig:footprint}
\end{figure}

The aim of the DESI BGS is to create a highly complete flux limited catalogue
of bright, low redshift galaxies, for the primary science goals of BAO and RSD analysis.
The survey is expected to cover $\sim 14,000$ square degrees (Fig.~\ref{fig:footprint})
in 3 passes of the sky, measuring 
spectroscopic redshifts of $\sim 10$ million galaxies, approximately 2 magnitudes
deeper than the SDSS main survey \citep{Strauss2002}, 
with double the median redshift ($z_{\rm med} \sim 0.2$).
The BGS will take place concurrently with the Milky Way 
Survey during bright time, when the sky is too bright for the main dark time survey
due to moon phase and twilight conditions. 

Fibres are currently planned to be assigned to science targets based on the following priority tiers:
\begin{enumerate}
\item Priority 1 galaxies ($r<r_\mathrm{bright}$, $\sim 800$ deg$^{-2}$) 
\item Priority 2 galaxies ($r_\mathrm{bright}<r<r_\mathrm{faint}$, $\sim 600$ deg$^{-2}$) 
\item Milky Way stars
\end{enumerate}
where $r_\mathrm{bright} \sim 19.5$ and $r_\mathrm{faint} \sim 20.0$.\footnote{In Section~\ref{sec:surveysims}
we use $r_\mathrm{bright}=19.452$ and $r_\mathrm{faint}=19.925$, which in the BGS mock
catalogue gives number densities of 818~$\persqdeg$ and 618~$\persqdeg$ for the bright
and faint samples respectively. We also randomly promote 10\% of the faint sample to
have the same priority as the bright sample (see Section~\ref{sec:fib_assign}).} 

The brightest galaxies with an $r$-band magnitude $r<19.5$ are preferentially targeted, since
the redshift success rate is expected to be high, making this sample of galaxies highly complete. 
Fainter galaxies, with $19.5<r<20.0$, which have a lower redshift success rate, are given a lower priority, 
and will form a less complete sample. 
If a fibre cannot be placed on a galaxy, it will be placed on a Milky Way star.

If a galaxy fails to have its redshift measured, one possibility is for it to remain at the same priority
in the next pass. If a redshift is successfully measured, it will remain a potential
target in future passes to give the possibility of improving the signal-to-noise
of the spectra, but its priority demoted to a fourth priority tier (below that of the Milky Way stars).

In addition to the galaxy targets, 100 fibres will be positioned on standard stars and 
400 on blank sky locations (sky fibres) in each exposure, with an equal number per petal 
\red{(see Section~\ref{sec:petals})},
for flux calibration and sky subtraction.

The observation strategy that will be used in the BGS is still to be chosen. We assume
a strategy in which the 3 complete passes of the entire survey are observed sequentially. 
Each pass consists of
$\sim 2000$ tiles positioned over the entire survey footprint, with overlaps between
neighbouring tiles. In the first pass, the tile centres are positioned on the sky with
an icosahedral tiling. The tiling for subsequent passes is identical, except with a
rotation on the sky, which fills in the missing area due to gaps in
the focal plane. %(docDB-717\footnote{\url{https://desi.lbl.gov/DocDB/cgi-bin/private/ShowDocument?docid=717}}).
The percentage of the survey footprint
that is covered by $N$ overlapping tiles after each full pass, and also after 90\% of the first 
pass\footnote{90\% of 1 pass is chosen as a realistically 
incomplete dataset, representing what might be available one third of the way through the survey,
where certain fields are missed due to observing conditions.},
is summarised in Table~\ref{tab:overlaps}.
After 1 pass, $\sim 90\%$ of the footprint is covered by a single tile. This
is greatly reduced after subsequent passes, with $\sim 80\%$ covered by 3 or more
tiles at the end of the survey. These numbers take into account the gaps in the focal plane.

\begin{table} 
\caption{Percentage of the survey area covered by $N$ overlapping tiles after
1 pass with 10\% of tiles missing, and after the full 1, 2 and 3 passes.
The total area covered by each pass is calculated by finding the fraction of
objects in a random catalogue that can be potentially assigned a fibre.}
\begin{tabular}{ccccc}
\hline
$N$ & Pass 1 (90\%)   & Pass 1          & Pass 2         & Pass 3 \\
    & (12.2k deg$^2$) & (13.5k deg$^2$) & (14.6k deg$^2$)& (14.8k deg$^2$) \\
\hline
1 & 89.8  & 88.4  & 13.4   &  3.6   \\
2 & 10.2  & 11.6  & 67.3   & 14.9   \\
3 &  0.01 &  0.01 & 18.4   & 55.9   \\
4 &  0.0  &  0.0  &  0.9   & 23.1   \\
5 &  0.0  &  0.0  &  0.009 &  2.3   \\
6 &  0.0  &  0.0  &  0.0   &  0.1  \\
7 &  0.0  &  0.0  &  0.0   &  0.006\\
\hline
\end{tabular}
\label{tab:overlaps}
\end{table}

\subsection{Robotic Fibre Positioners}
\label{sec:petals}

\begin{figure} 
\centering
\includegraphics[width=\linewidth]{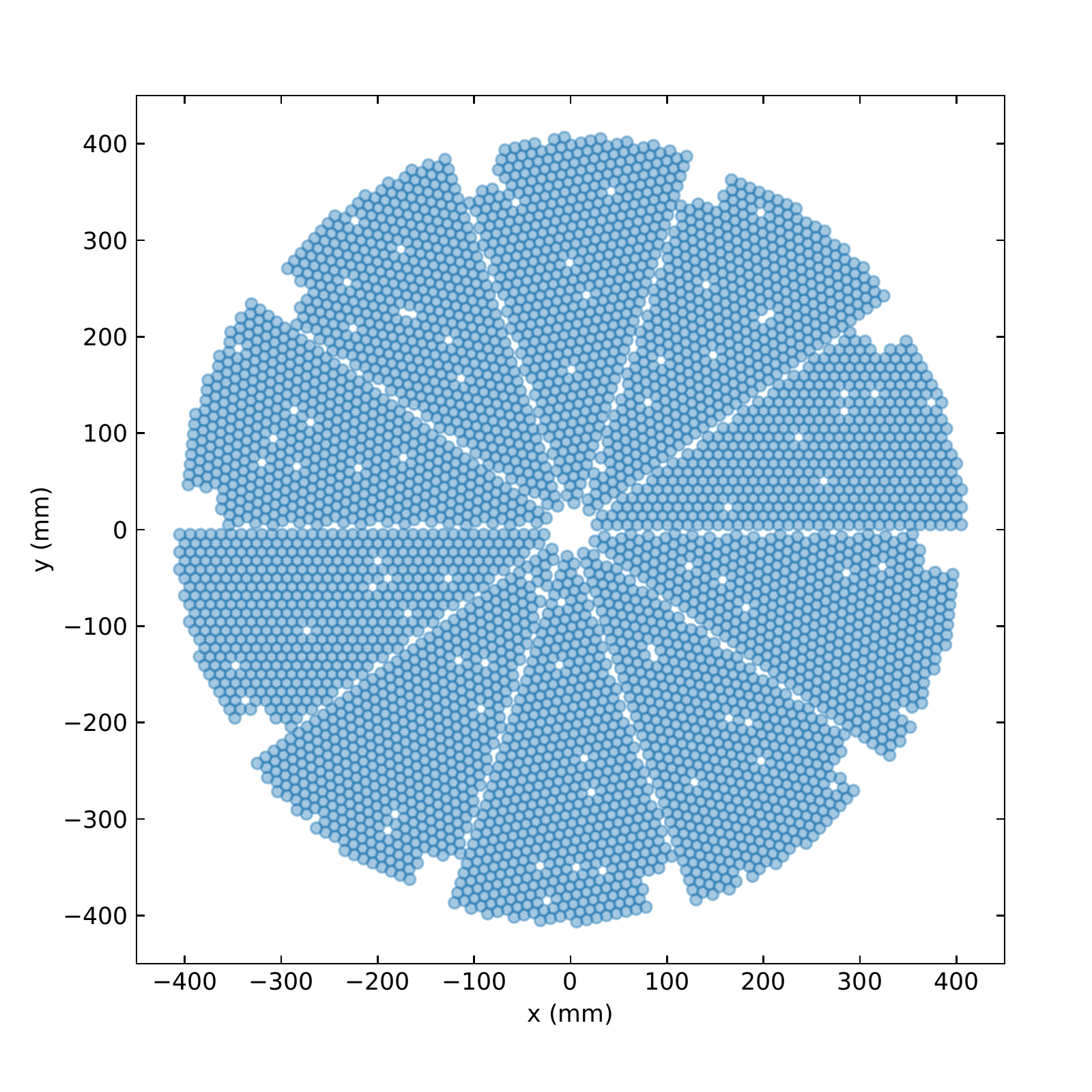}
\vspace{-0.5cm}
\caption{A single DESI tile, showing the arrangement of fibres in the focal plane,
split into 10 petals. The blue circles indicate the patrol area of each fibre. The holes
within each petal are the locations of the fiducials, which provide a light source for the 
fibre view camera to calibrate the placement of the fibre positioners.}
\label{fig:tile}
\end{figure}

Each pointing of DESI, or tile, consists of a total of 5,000 fibres, arranged on the
focal plane in 10 wedge-shaped `petals' \citep{Schubnell2016}. Each individual fibre is 
controlled by a 
robotic fibre positioner which can rotate on two arms, allowing the fibre to be placed 
on any object within a unique 
circular patrol region \citep[see e.g. figure~3.11 of][]{DESI2016b}, with
a patrol radius corresponding to an angle on the sky of 
$R_\mathrm{patrol} = 1.48$~arcmin (0.0247~deg) (at $z=0.2$, this is a comoving
separation of $0.25~\hMpc$). 
The arrangement of fibres is illustrated in Fig.~\ref{fig:tile}.
There is a small overlap between the patrol regions of neighbouring fibres,
and there are gaps between petals which cannot be reached by a fibre. The `missing' 
squares around the edge of the tile are the location of the guide focus arrays, \red{which measure the pointing of the telescope and
orientation of the focal surface}. Each petal also
contains 10 fiducials which provide light sources for the fibre view camera to calibrate 
fibre positioner placement \citep[section~3.5 of][]{DESI2016b}.

\subsection{Fibre Assignment Algorithm}
\label{sec:fibassign}

To assign fibres to targets, each potential target object is first assigned a primary priority,
which is an integer that is determined by the priority tier of the object, 
e.g. all priority 1 galaxies have the same primary priority, which is \red{numerically}
greater than the priority
2 galaxies. A uniform random sub-priority in the range $(0,1)$ is then
generated for each object, and the total priority is the sum of the primary and sub-priorities.
Fibres are ordered by the highest priority object in their patrol region (from highest to lowest \red{numerical value}),
and are looped through in this order. Each fibre in turn is assigned to the object in its patrol
region with the highest priority it is possible for it to target. With this scheme,
the assignment of fibres to objects in the same priority tier is randomized, 
but if a high priority object competes for a fibre with a low priority object, the high 
priority object will always be assigned a fibre at the expense of the low priority object.
If fibres are instead looped through in a fixed order, certain fibres would always have a 
high priority, and be assigned to a galaxy before its six neighbouring fibres,
potentially preventing them from ever targeting certain objects due 
to fibre collisions.

In the current survey strategy, the entire survey is split into several epochs. In each epoch, tiles are selected by a survey planning algorithm, which determines the sequence of tiles based on date and survey conditions. The selected tiles then go through the fibre assignment algorithm. The fibre assignment algorithm loops through each tile, in a fixed order, assigning fibres to objects. At the end of this loop, there is some redistribution of fibres so that
\begin{enumerate}
\item the total number of targets observed is maximized
\item there are the required number of standard stars and sky fibres
\item fibres that are unused are uniformly distributed over tiles.
\end{enumerate}
After fibre assignment, at the end of the epoch, galaxy priorities are updated depending on whether the redshift measurement was successful or unsuccessful. The updated galaxy priorities is then used in the next epoch. 
%(docDB-2742\footnote{\url{https://desi.lbl.gov/DocDB/cgi-bin/private/ShowDocument?docid=2742}}). 

In order to make unbiased 2-point galaxy clustering measurements using the \citet{Bianchi2017} scheme, each pair of 
objects in the parent sample must have a non-zero probability of being targeted (see Section~\ref{sec:pip}).
To make sure as many pairs as possible can be targeted, we do the following:

\subsubsection{Dithering tile positions}

In regions covered by a single tile, if there are two priority 1 galaxies in the
unique patrol region of a single fibre, that fibre will target the galaxy with the 
\red{numerically} highest
random sub-priority, but it can never target both, so the pair will always be missed.

\red{This can be mitigated by, in each realization of fibre assignment,
applying a global dither to the tiling of the entire survey, i.e.
randomly rotating the whole 3-pass set of survey tiles by a small angle
(of the order of $R_\mathrm{patrol}$). This is entirely equivalent
to keeping the tiling fixed in each realization, and rotating the
galaxy positions on the celestial sphere}.
%This can be mitigated by dithering the tile positions in each realization of the fibre
%assignment algorithm, i.e. randomly rotating the whole set of survey tiles by a small angle
%(of the order of $R_\mathrm{patrol}$). 
In some
of these random dithers, the two objects in an untargetable pair will be split
between two neighbouring fibres, giving the pair a non-zero probability of being
targeted. Since tile centres are uncorrelated with large scale structure, galaxy pairs of any
separation in any environment are equally likely to be targeted in each realization, and
therefore it is valid to average over realizations to estimate the probability.
To dither
the tile positions, a random rotation axis is chosen, which is uniformly distributed.
The tile centres are then rotated around this axis by a small angle, which we 
choose to be 3 times the fibre patrol radius.

The dithering of the tile positions is only done when applying the pair weighting correction
described in Section~\ref{sec:pip}. When assigning fibres to objects in the real survey, the
rotation angle is set to zero.

\subsubsection{Priority 2 galaxies}

Priority 1 galaxies always have a higher priority than priority 2 galaxies, so if it is
possible for a fibre to be placed on an unobserved priority 1 galaxy, it will always target that galaxy,
regardless of how many priority 2 galaxies are in the same patrol region. 
This means that a significant fraction of priority 2 galaxies in regions 
with a high density of priority 1 galaxies will always be missed.

One way of sampling these missing priority 2 galaxies is, in each fibre assignment realization,
to randomly promote a certain fraction of priority 2 galaxies to the same priority as 
the priority 1 galaxies. This gives pairs containing at least one priority 2
galaxy in over-dense regions
a small, but non-zero probability of being targeted (see Fig.~\ref{fig:fibre_targets}).

The version of the fibre assignment algorithm we use is 0.6.0.\footnote{\url{https://github.com/desihub/desitarget}}

\subsection{Survey Simulations}
\label{sec:surveysims}

To quantify incompleteness due to fibre assignment and assess correlation function
correction methods, we run the fibre assignment algorithm on a BGS mock
catalogue from the Millennium-XXL (MXXL) simulation \citep{Smith2017}. This is 
a halo occupation distribution (HOD) mock, which contains galaxies to $r=20$
over the same redshift range as the BGS, and is constructed to reproduce the
luminosity function and clustering measurements from SDSS \citep{Blanton2003,Zehavi2011} and 
GAMA \citep{Loveday2012,Farrow2015}.\footnote{The MXXL mock is available at \url{http://icc.dur.ac.uk/data/} 
and \url{https://tao.asvo.org.au/tao/}}

The magnitudes in this catalogue are in the SDSS $r$-band. These are converted to
the DECam $r$-band (which is used in the DESI target selection) using
\begin{equation}
r_\mathrm{DECam} = r_\mathrm{SDSS} - 0.03587 - 0.14144(r-i)_\mathrm{SDSS}.
\end{equation}
%(docDB-1788\footnote{\url{https://desi.lbl.gov/DocDB/cgi-bin/private/ShowDocument?docid=1788}}). 
Since the mock catalogue does not contain $r-i$ colours, we assume a mean colour of
$(r-i)=0.4$. To make sure the priority 1 and 2 galaxies have number densities 
of 818 deg$^{-2}$ and 618 deg$^{-2}$, we define priority 1 and 2 galaxies using
the magnitudes $r_\mathrm{DECam}=19.452$ and $r_\mathrm{DECam}=19.925$.\footnote{These number 
densities are chosen to match assumptions made in
earlier survey simulations (J. Tinker, private communication).}

The mock is first cut to the set of galaxies which are within the patrol
radius of a fibre in the full 3-pass survey (with no dither), and therefore could potentially be
assigned a fibre.\footnote{In our clustering analysis we account for the regions of sky this process discards by applying the same criterion to the corresponding random catalogue. This differs from
\citet{Bianchi2018}, in which the random sample covers the full survey volume.}
We run 2048 random realizations of the
fibre assignment algorithm ($\sim 500$ CPU hours), with the full 3 passes of tiles to simulate the complete
survey. From the survey simulation output, it is also possible to determine
which galaxies were assigned fibres in the first or second pass, allowing
us to simulate a more incomplete survey without having to re-run the
fibre assignment code. In addition to the full 3 passes, we also determine
which galaxies are targeted in
1 pass, with a random 10\% of tiles missing (which are the same tiles in each realization),
to simulate a dataset that might realistically be achieved after 1/3 of the duration of the survey
with a survey strategy that prioritizes area (i.e. a strategy where after 1/3 of the duration,
pass 1 is completed, as opposed to a strategy where 3 passes are completed in only 1/3 of the
survey area).
Removing tiles reduces the overall area of the footprint and increases the fraction of 
the remaining area that is covered by a single tile.

In each run of the fibre assignment code, the tile positions are randomly dithered
by an angle 3 times the patrol radius, and a random 10\% of priority 2 galaxies are 
promoted to the same priority as the bright sub-sample. Unless
specified, we will refer to the bright sub-sample as `priority 1' and the faint
sub-sample as `priority 2'.

%We assume that if a fibre is placed on a galaxy, a redshift measurement is
%successful, and do not simulate the effects of weather, in order to focus solely on
%incompleteness caused by the fibre assignment algorithm.
We only consider targeting incompleteness caused by the fibre assignment algorithm.
Redshift incompleteness due to redshift measurement failures, and the effects of 
weather, are left for future work.

\section{Fibre Assignment Completeness}
\label{sec:completeness}

\begin{figure*}
\centering
\includegraphics[width=\linewidth]{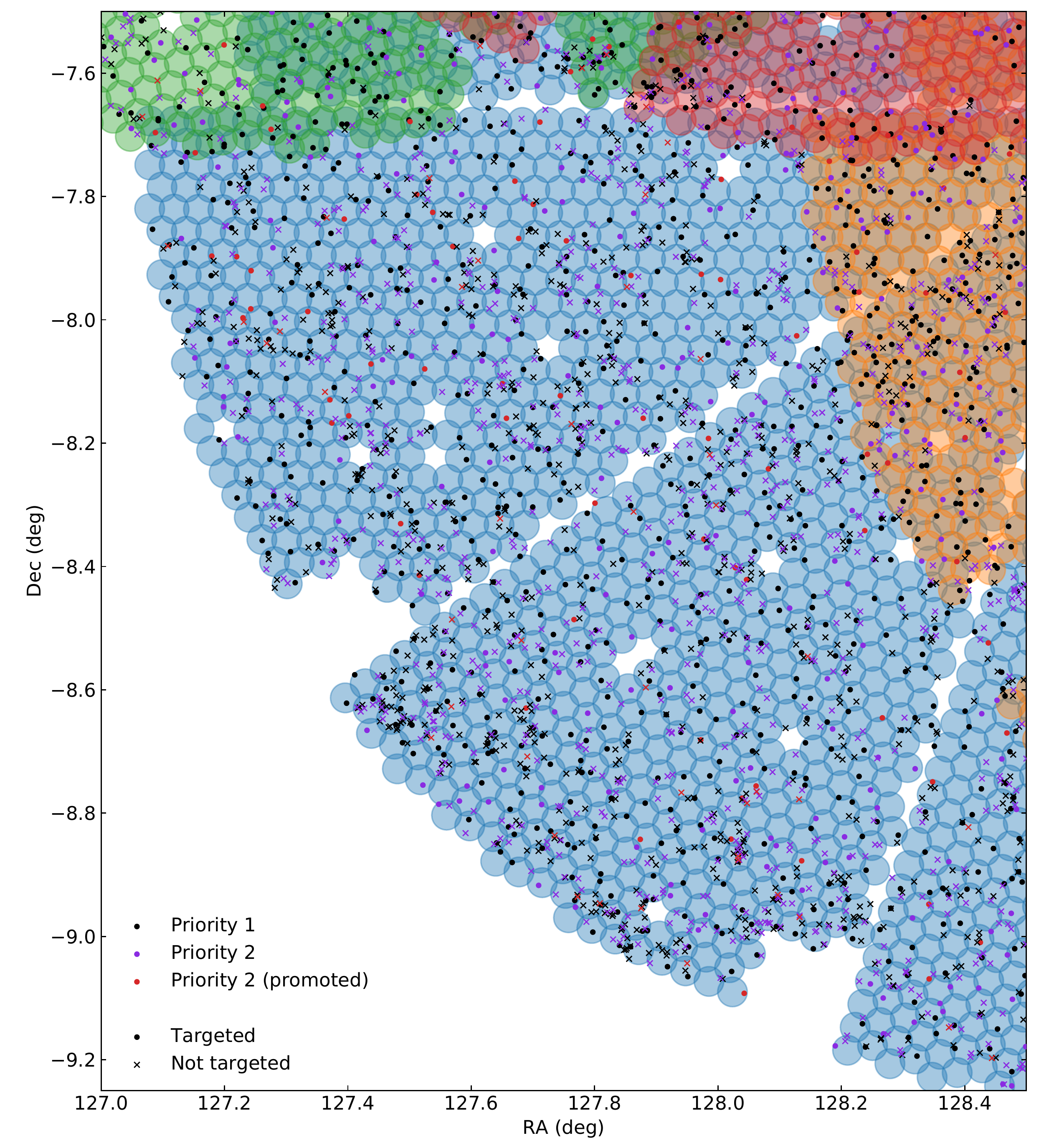}
\caption{A zoom in on a small section around the edge of the survey footprint of
one survey simulation, showing the positions of BGS galaxies relative to fibre patrol
regions. This survey simulation has zero dither, but 10\% of priority 2 galaxies are
randomly promoted. Shaded circles indicate the patrol region of each fibre, with
each neighbouring tile in a different colour. White regions cannot be reached by a fibre.
Circles indicate galaxies which are successfully assigned a fibre, while
crosses show untargeted galaxies. The bright priority 1 sample is shown in
black, and the faint priority 2 sample is in purple. Promoted priority 2 galaxies are
shown in red.}
\label{fig:fibre_targets}
\end{figure*}

For a small region of sky, Fig.~\ref{fig:fibre_targets} shows the positions of targeted and 
untargeted galaxies in the BGS mock with the fibre patrol regions superimposed. This region is 
at the edge of the survey, and is mostly covered by a single tile, shown in blue, with neighbouring tiles 
in different colours.
%relative to the fibre patrol regions (with no dither),
%indicating the galaxies that are successfully assigned fibres in one survey simulation.
On the scale of the fibre patrol regions, the surface density of galaxies varies greatly. 
Some fibres have zero galaxies in their
patrol region, leaving them free to target Milky Way stars, while fibres in
dense regions can have 10 or more galaxies within their patrol region. It is 
clear to see that in dense regions, the fibre assignment completeness will be
low, since only one galaxy can be assigned a fibre out of many potential targets.
More galaxies can be targeted if there are multiple tile overlaps, which will make
the completeness higher.
In low density regions, the completeness will be very high, 
since if there is only 1 galaxy within a fibre patrol region,
the fibre will always be placed on that galaxy.

\begin{figure} 
\centering
\includegraphics[width=\linewidth]{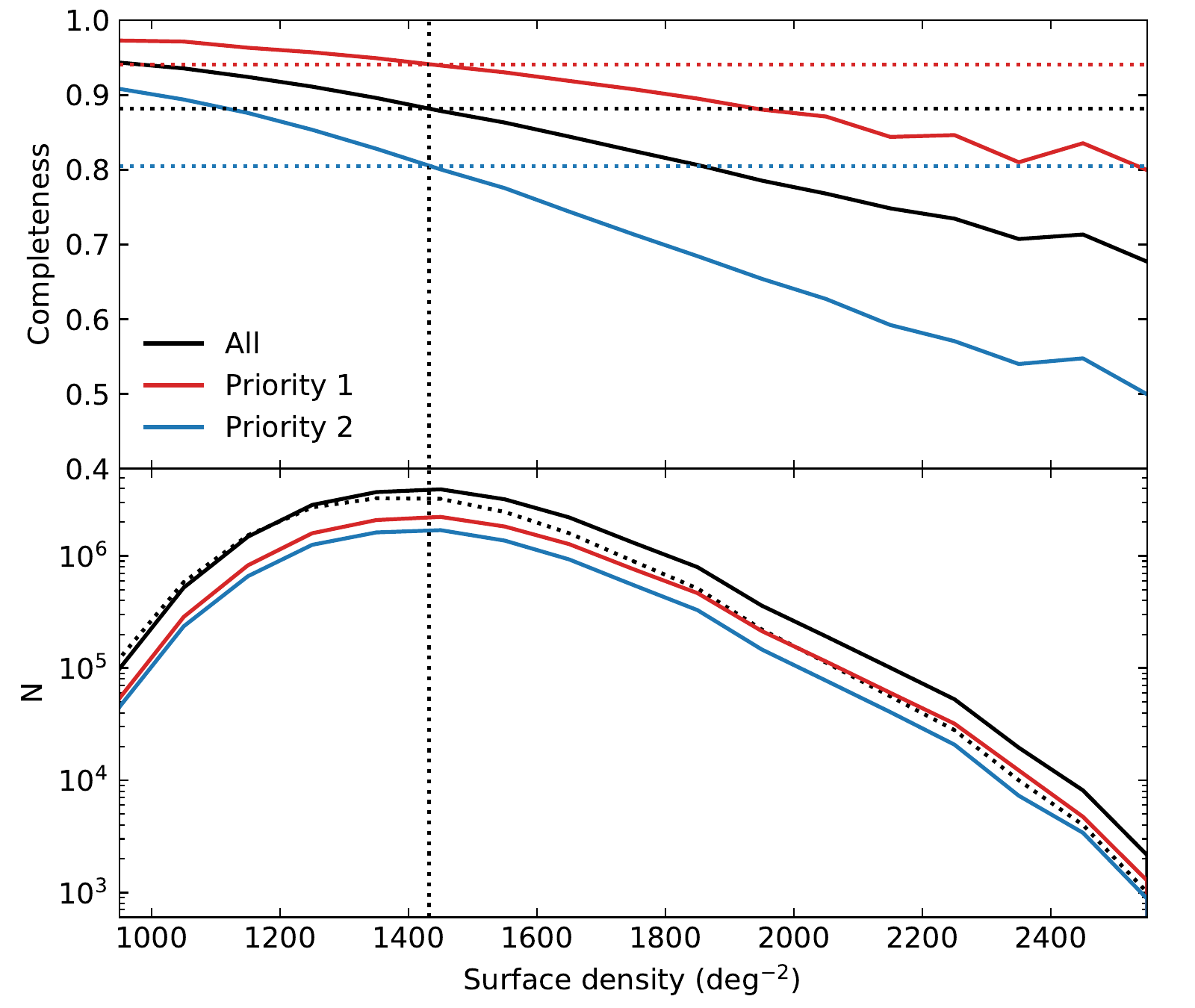}
\caption{\textit{Top panel:} average fibre assignment completeness as a function of 
the surface density of all BGS galaxies,
in \textsc{healpix} pixels of area 0.84 deg$^2$ ($N_\mathrm{side}=64$) for all galaxies (black),
priority 1 (red) and priority 2 (blue), after 3 passes with 10\% of priority 2 galaxies promoted. 
The vertical dotted line indicates the average
surface density of the survey ($1436~\persqdeg$), and horizontal dotted lines indicate the median completeness
for the three samples (88\%, 94\% and 80\% for all, priority 1 and priority 2 galaxies respectively).
\textit{Bottom panel:} histogram of the total number of objects in bins of surface density. The dotted black curve
shows the number of \textsc{healpix} pixels, scaled up by a factor of 1000.}
\label{fig:completeness_vs_density}
\end{figure}

The completeness due to surface density is quantified in Fig.~\ref{fig:completeness_vs_density}.
The upper panel shows the average completeness 
as a function of surface density, after 3 passes, in \textsc{healpix} pixels \citep{Gorski2005} 
with area 0.84 $\sqdeg$ ($N_\mathrm{side}=64$), separately for all galaxies, and for priority 1 and 2 galaxies.
The completeness decreases monotonically as the surface density of galaxies
increases. Also, since priority 1 galaxies are preferentially targeted, they have a higher
completeness than the priority 2 galaxies. The vertical dotted line indicates a
surface density of $1436~\sqdeg$,
which is the average surface density of all (priority 1 and 2) galaxies, 
and horizontal dotted lines show the median completeness in \textsc{healpix} pixels, which is 88\%,
94\% and 80\% for all, priority 1, and priority 2 galaxies respectively.
The lower panel shows a histogram of the total
number of galaxies, which peaks close to the average surface density.
%\textbf{ JH: the mean density for priority 1 and priority 2 galaxies are 800 and 600, which can not be read from this plot. It seems that they all peak at 1400.} 
The black dotted curve shows the histogram of the densities of individual \textsc{healpix} pixels, 
scaled up by a factor of
1000. The unscaled black dotted curve, multiplied by the average number of galaxies per pixel,
produces the black solid curve. The variance in the surface density of pixels depends on the
resolution. For pixels with area 13.4 $\sqdeg$ 
($N_\mathrm{side}=16$), which is larger than the area of each tile, the surface density varies from
the mean by a few hundred objects per square degree.

The fibre assignment completeness of galaxies in the BGS is driven by the surface density of 
galaxies, since 
it is not possible to place a fibre on every galaxy if the density of galaxies is greater than
the density of fibres.\footnote{Each tile of 5000 fibres has a radius of 1.605 deg,
which corresponds to a fibre surface density of $\sim 600~\persqdeg$.} 
With multiple passes, the same area of sky will be re-observed
several times, enabling some of these previously missed galaxies to be targeted. After
the full 3 passes of the BGS, most of the footprint ($\sim 80\%$) will have been covered 
by 3 or 4 tiles (see Table~\ref{tab:overlaps}), but the targeted catalogue will still be incomplete 
in high density regions.

\begin{figure} 
\centering
\includegraphics[width=\linewidth]{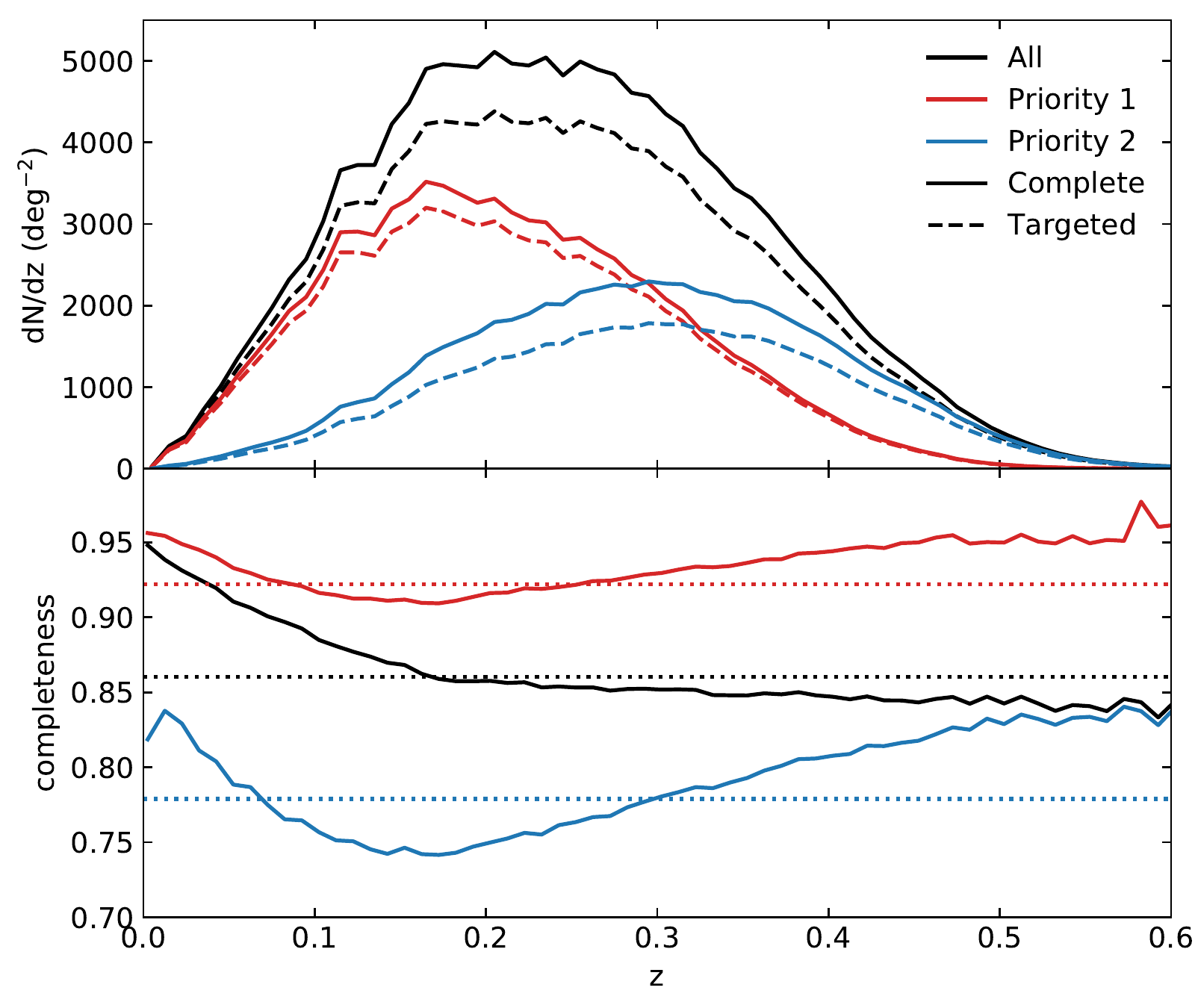}
\caption{\textit{Top panel:} Redshift distribution of galaxies before and after fibre 
assignment (solid and dashed curves), with the full 3 passes of tiles. 
The complete sample of BGS galaxies is shown in black,
while priority 1 and priority 2 galaxies are in red and blue respectively.
\textit{Bottom panel:} Completeness as a function of redshift for all, priority 1 and
priority 2 galaxies. Horizontal dotted lines indicate the mean completeness 
(86\%, 92\% and 78\% for all, priority 1, and priority 2 galaxies respectively).}
\label{fig:completeness_vs_redshift}
\end{figure}

The upper panel of Fig.~\ref{fig:completeness_vs_redshift} shows the redshift distribution
of galaxies in the BGS, before and after fibre assignment (solid and dashed curves). The lower 
panel shows the targeting completeness as a function of redshift, 
where the horizontal dotted lines indicate the average
completeness. For the priority 1 and the priority 2 galaxies, this curve is non-monotonic.
This is because haloes at high redshifts contain few galaxies brighter than the magnitude limit.
These galaxies will not greatly enhance the surface density, and the completeness is high. At
intermediate redshifts, many more galaxies per halo can be detected in haloes of the same mass, which
will result in a much greater enhancement of the surface density, and therefore a lower completeness.
At low redshifts, haloes of the same mass will contain an even greater number of 
galaxies brighter than the magnitude
limit, but since they are nearby, they subtend a relatively large angle on the sky, and the
perturbation to the surface density is low again. For the complete galaxy sample, the completeness
is relatively flat at high redshifts, since the fraction of priority 2 galaxies increases
with redshift.

The mean completeness (which differs slightly from the median completeness shown
in Fig.~\ref{fig:completeness_vs_density}) is $\sim 86\%$, while for 
priority 1 and 2 galaxies it is $\sim 92\%$ 
and $\sim 78\%$ respectively. These figures are for the case where 10\% of the priority 2 galaxies
are given the same priority as the priority 1 galaxies. If there was no promotion of priority 2 
objects, the priority 1 galaxies would be more complete, ($\sim 93 \%$) but at the expense of the 
low priority galaxies (see Table~\ref{tab:promotion}).

\begin{figure} 
\centering
\includegraphics[width=\linewidth]{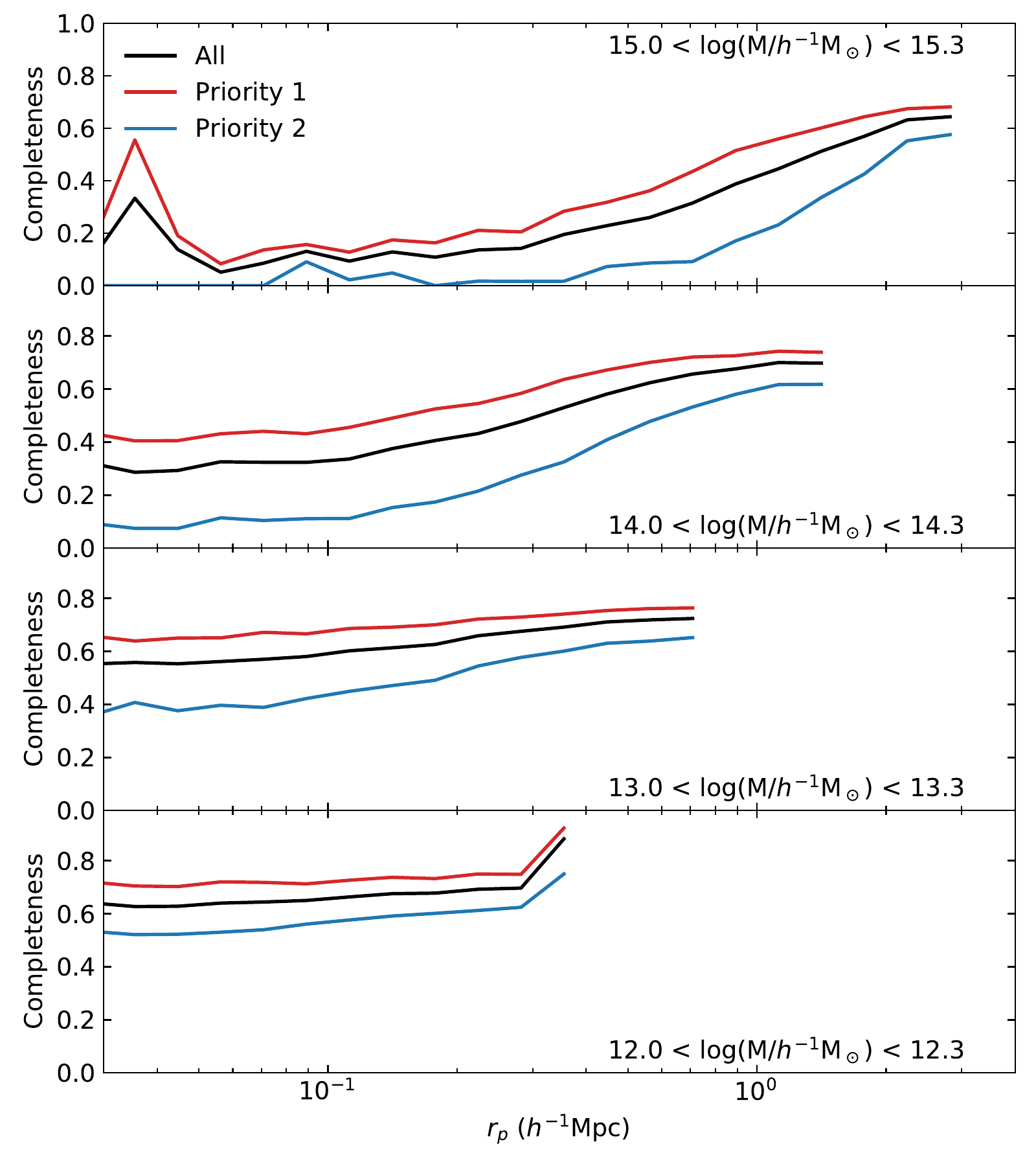}
\caption{Targeting completeness of galaxies in haloes as a function of the transverse distance 
from the centre of their respective halo, for haloes in the redshift
range $0.15<z<0.25$, after 3 passes. The completeness for all galaxies is shown in black,
and for priority 1 and 2 galaxies in red and blue respectively.}
\label{fig:completeness_vs_halo_distance}
\end{figure}

Fig.~\ref{fig:completeness_vs_halo_distance} shows the completeness of galaxies in haloes,
as a function of the distance from the centre of their host halo, for haloes in different mass bins
around the peak of the redshift distribution ($0.15<z<0.25$). The panels, from top to bottom, show 
the completeness
for haloes with masses $M_\mathrm{200mean} \sim 10^{15} \hMpc$, $M_\mathrm{200mean}\sim 10^{14} \hMpc$, 
$M_\mathrm{200mean} \sim 10^{13} \hMpc$, and $M_\mathrm{200mean} \sim 10^{12} \hMpc$
respectively, plotted to the virial radius ($R_\mathrm{200mean}$). 
$M_\mathrm{200mean}$ is defined as the mass enclosed by a sphere of radius $R_\mathrm{200mean}$,
in which the average density is 200 times the mean density of the Universe.
Close to the centre of large haloes, the surface density of galaxies is very high,
and therefore the completeness is very low. For $10^{12} \hMpc$ haloes, the average completeness 
near the centre is $~\sim 60\%$, but for the most massive haloes, this completeness is much
lower. The spike close to the centre of $M \sim 10^{15} \hMpc$ haloes is due to 
noise. When measuring two-point clustering statistics, as we show in Section~\ref{sec:results}, the
effect of this incompleteness can be corrected, and this is unbiased so long as each galaxy pair 
has a non-zero probability of being targeted.
Since the completeness in clusters is low, care must be taken, for example, identifying clusters
and voids and estimating velocity dispersions. The incompleteness must also be taken into 
account when estimating higher-point statistics. Our realizations of the fibre assignment
algorithm could be used to develop correction procedures for these statistics.

The total number of objects targeted, and the completeness after each pass, 
is shown in Table~\ref{tab:completeness} for all galaxies, priority 1 and 2
galaxies, and the subset of priority 2 galaxies that are promoted to the
same priority as priority 1. Since faint galaxies are less clustered than
bright galaxies, the promoted priority 2 galaxies have a higher completeness
than the priority 1 galaxies. Most of the promoted galaxies are targeted in
the first pass.

Table~\ref{tab:promotion} shows how the final completeness after 3 passes is
affected by the fraction of objects in the faint sub-sample promoted to
high priority. The priority 1 sample is most complete with zero promotion
(92.9\%), but the priority 2 sample is least complete (77\%), and certain
priority 2 objects will always be missed due to conflicts with high
priority objects. As the fraction of priority 2 objects is increased, the
percentages converge to the average completeness of $\sim 86\%$. 

\begin{table}
\caption{Table showing the cumulative number of objects targeted after each pass,
in millions, and the completeness, as a percentage. Priority 1 and priority 2
are the intrinsic priorities based on magnitude. Priority 2 (p) is the subset
of priority 2 galaxies that are promoted to have the same priority as the
bright priority 1 galaxies. The final row shows the cumulative number of unused fibres
which are available to target Milky Way stars (in millions) after each pass, 
and the percentage of fibres which are unused after each pass.
A total of $\sim 9$ million fibres are available
per pass, excluding standard stars and sky fibres (2,000 pointings, 
each with 4,500 available fibres).}
\begin{tabular}{ccccccc}
\hline
\multirow{2}{*}{Sample} & \multicolumn{2}{c}{Pass 1} & %
    \multicolumn{2}{c}{Pass 2} & \multicolumn{2}{c}{Pass 3}\\
 & $N_\textrm{gal}$ & \% & $N_\textrm{gal}$ & \% & $N_\textrm{gal}$ & \% \\
\hline
All            & 7.54 & 35.6 & 13.78 & 65.0 & 18.24 & 86.0 \\
Priority 1     & 5.15 & 42.7 &  8.84 & 73.3 & 11.11 & 92.2 \\
Priority 2     & 2.39 & 26.1 &  4.95 & 54.1 &  7.12 & 77.8 \\
Priority 2 (p) & 0.79 & 86.2 &  0.84 & 92.4 &  0.85 & 93.2 \\
\hline
Free fibres    & 1.49 & 16.5 &  4.30 & 23.8 &  8.89 & 32.8 \\
\hline
\end{tabular}
\label{tab:completeness}
\end{table}

\begin{table}
\caption{Table showing the number of objects targeted after 3 passes, in millions, and the 
completeness, in survey simulations where the percentage of promoted priority 2 galaxies is varied 
from 0\% to 40\%. Priority 1 and 2 galaxies are the bright and faint sub-samples, and priority 2 (p)
are the promoted subset of priority 2 galaxies.}
\begin{tabular}{ccccccc}
\hline
\multicolumn{1}{c}{Promotion} & \multicolumn{2}{c}{Priority 1} & %
    \multicolumn{2}{c}{Priority 2} & \multicolumn{2}{c}{Priority 2 (p)}\\
\% & $N_\textrm{gal}$ & \% & $N_\textrm{gal}$ & \% & $N_\textrm{gal}$ & \% \\
\hline
 0 & 11.12 & 92.9 & 7.04 & 77.0 & -    & -    \\
 5 & 11.15 & 92.5 & 7.08 & 77.4 & 0.43 & 93.7 \\
10 & 11.11 & 92.2 & 7.12 & 77.8 & 0.85 & 93.2 \\
15 & 11.07 & 91.8 & 7.17 & 78.4 & 1.28 & 93.1 \\
20 & 11.02 & 91.5 & 7.21 & 78.9 & 1.69 & 92.6 \\
25 & 11.00 & 91.1 & 7.26 & 79.3 & 2.11 & 92.5 \\
30 & 10.94 & 90.7 & 7.30 & 79.8 & 2.52 & 92.0 \\
35 & 10.89 & 90.3 & 7.35 & 80.3 & 2.93 & 91.7 \\
40 & 10.84 & 90.0 & 7.39 & 80.8 & 3.34 & 91.4 \\
\hline
\end{tabular}
\label{tab:promotion}
\end{table}

\section{Correcting Two-Point Clustering Measurements}
\label{sec:corrections}

\subsection{Mitigation Techniques}

The two-point correlation function at separation $\vec{s}$ can be estimated using the 
estimator of \citet{Landy1993},
\begin{equation}
\xi(\vec{s}) = \frac{DD(\vec{s}) - 2DR(\vec{s}) + RR(\vec{s})}{RR(\vec{s})},
\end{equation}
where $DD$, $DR$ and $RR$ are the normalized data-data, data-random, and 
random-random pair counts. If galaxies in the data catalogue are missing,
the resulting correlation function will be biased. Mitigation techniques 
attempt to recover the correlation function of the parent sample from the
sample of galaxies that are targeted.

\subsubsection{Nearest object}
We use two different nearest redshift corrections. In the first correction,
missing galaxies are assigned the redshift of the nearest targeted object
on the sky \citep[the approach taken in the SDSS survey analyses in e.g.][]{Zehavi2005, Berlind2006, Zehavi2011}.
The catalogue of galaxies is then cut to the volume limited
sample using these redshifts. Some of the untargeted objects will be 
assigned a redshift close to the true value, and will be correctly 
identified as part of the volume limited sample, but the sample
will be contaminated by other galaxies which are assigned incorrect
redshifts. We refer to this correction as `nearest redshift'.

In the second correction, each galaxy is first given a weight of 1, and the
weight of a missing galaxy is added to the nearest targeted object on the sky 
\citep[e.g. in BAO analysis in the BOSS survey,][]{Anderson2012, Anderson2014a, Anderson2014b}.
For example, a targeted
galaxy with no nearby untargeted galaxies would have weight 1. If there
was a close galaxy that was not targeted, the weight would be transferred
to the targeted galaxy, which would now have a weight of 2. We hereafter
refer to this correction as `nearest weight'.
The nearest weight correction can be seen as an approximation of the pair
weighting method of Section~\ref{sec:pip} \citep[see][]{Bianchi2017}.

\subsubsection{Angular upweighting}
When estimating the correlation function, galaxy pairs are upweighted by the factor
\begin{equation}
W(\theta) = \frac{1 + w^{(p)}(\theta)}{1 + w(\theta)},
\end{equation}
where $w^{(p)}(\theta)$ is the angular correlation function of the complete, parent sample of
galaxies, and $w(\theta)$ is the incomplete, targeted sample~\citep[e.g. the 2dFGRS analysis of][]{Hawkins2003}. 
This angular weighting by construction recovers the angular correlation of the 
parent sample. This correction makes the assumption that the targeted and untargeted
galaxies are statistically equivalent in each angular bin, which is not necessarily true, and therefore
it may not provide an adequate correction to the redshift space correlation function.

\subsubsection{Pair Inverse Probability (PIP) Weights}
\label{sec:pip}

The PIP weighting scheme \citep{Bianchi2017} upweights each galaxy pair by the pair weight
$w_{ij} = 1 / p_{ij}$, where $p_{ij}$ is the probability that the pair will
be targeted. This probability can be estimated by running the fibre assignment
code $N_\mathrm{real}$ times, where $N_\mathrm{real}$ is of the order of 100s or 
1000s. For galaxy $i$, a vector $\vec{w}_i$ of length $N_\mathrm{real}$ is 
stored, which contains a 1 if the galaxy is assigned a fibre, and a 0 otherwise.
This vector can conveniently be stored as the bits of an integer (or several integers).
The pair weight for galaxies $i$ and $j$ can be written as the dot-product of
these vectors, but can be efficiently calculated using bitwise operations,

\begin{equation}
w_{ij} = \frac{N_\mathrm{real}}{\vec{w}_i \cdot \vec{w}_j} \equiv \frac{N_\mathrm{real}}{\mathrm{popcount}(\vec{w}_i \& \vec{w}_j)},
\label{eq:pair_weights}
\end{equation}
where \& is the bitwise `and' operator, and popcount is a bitwise operator which
sums together the bits of an integer.

The corrected DD counts are calculated from summing the pair weights of galaxies
in the separation bin $\vec{s}$,
\begin{equation}
DD_w(\vec{s}) = \sum_{\vec{s}_i-\vec{s}_j \approx \vec{s}} w_{ij} \frac{DD^{(p)}(\theta_{ij})}{DD_w(\theta_{ij})},
\label{eq:dd_pip}
\end{equation}
where $DD^{(p)}(\theta_{ij})$ are the angular DD counts of the parent sample,
and $DD_w(\theta_{ij})$ are the angular DD counts of the targeted sample but weighted by
the pair weights $w_{ij}$ (from Eq.~\ref{eq:pair_weights}), i.e.
\begin{equation}
DD_w(\theta) = \sum_{\Delta \theta_{ij} \approx \theta} w_{ij}.
\end{equation}
A similar correction is also applied to the DR counts, but this can be done
using individual galaxy weights (see Section~\ref{sec:iip}),
\begin{equation}
DR_w(\vec{s}) = \sum_{\vec{s}_i-\vec{s}_j \approx \vec{s}} w_{i} \frac{DR^{(p)}(\theta_{ij})}{DR_w(\theta_{ij})}.
\label{eq:dr_pip}
\end{equation}
In the case where there are no untargetable pairs the PIP estimator is unbiased\footnote{Pair 
weighting takes into account correlations between galaxies in a pair, 
and is unbiased if each pair has a non-zero probability of being
targeted. E.g. if a pair is targeted $n$ times in $N_\mathrm{real}$ fibre assignment realizations, its weight 
is $N_\mathrm{real}/n$, and it is targeted in $n/N_\mathrm{real}$ realizations, 
therefore the average weight is 1.}
without \red{the additional angular weighting factor in Eq.~\ref{eq:dd_pip}~\&~\ref{eq:dr_pip}. }
In this case the ensemble mean of the angular weighting factor is unity and its inclusion is to reduce the variance in the estimator \citep[see][]{Percival2017}. However, in the case where there are untargetable pairs, the PIP estimator without this factor is 
biased.\footnote{Note that since the pairs with zero probability never enter the pair counts, the
expectation value of the estimator is the clustering of the non-zero probability pairs.}
Including the angular weighting corrects this bias if, at any separation, the untargeted pairs are an unbiased sample of all the pairs of that separation. The accuracy of this assumption depends on the details of the targeting algorithm. Our results provide a direct test of this for the case of the DESI BGS.

\citet{Bianchi2018} apply the PIP weighting scheme to a DESI ELG mock catalogue,
and are able to recover unbiased clustering measurements. However, they do not
dither the tile positions, and rely entirely on the angular weighting term to
recover the small scale clustering. They also only include ELGs in their catalogue,
so do not consider objects with different priorities.

\subsubsection{Individual Inverse Probability (IIP) Weights}
\label{sec:iip}

Each galaxy is given an individual weight, which is the inverse of the probability
that the galaxy will be targeted, $w_i = 1 / p_i$. This can be estimated from
the same bitwise vectors used to estimate the pair probabilities,

\begin{equation}
w_i = \frac{N_\mathrm{real}}{\mathrm{popcount}(\vec{w}_i)}.
\end{equation}

%By running the fibre assignment
%algorithm many times, this weight can be estimated from 
%%$w_i = N_\textrm{real}/N_\textrm{targ}$, where $N_\textrm{real}$ is the number of
%realizations, and $N_\textrm{targ}$ is the number of realizations in which the galaxy
%is assigned a fibre.

If galaxies are given individual weights, the weight given to a pair of galaxies
is the product of these two weights, $w_{ij} = w_i w_j$. This pair weight does
not take into account any correlation between galaxy pairs, and will not
produce an adequate correction on small scales where pairs are highly
correlated.

\subsection{Clustering Estimates}

Correlation functions are calculated using the publicly available parallelized
correlation function code \textsc{twopcf}\footnote{\url{https://github.com/lstothert/two_pcf}},
which contains an efficient implementation of the PIP weighting scheme. The code can also
efficiently calculate jackknife errors in a single loop over the galaxy pairs
\citep{Stothert2018}.
To create the random catalogue, we uniformly generate random points on the sky, only keeping
those that fall within the patrol region of a fibre, with no dither, so that the random
catalogue covers the same footprint as the input catalogue. 
For illustrative purposes to compare correlation function correction techniques,
we assume the parent volume limited sample is known, and assign each object in the
random catalogue a redshift randomly sampled from this distribution.
This ensures that the number density of objects in the random catalogue
has the same evolution as the data catalogue.
In the real survey, the parent sample is not known beforehand, but the redshift distribution
can be determined by weighting the redshift distribution of the targeted sample
by the individual galaxy weights. We have checked, and the scatter between fibre assignment realizations
of the weighted $n(z)$ is within 1\%. Note that in the case 
of a flux limited catalogue, the parent sample is known, and
this is not an issue.

We also normalise the correlation function using the total number of objects in the
parent sample. Again, in the real survey, this is not known, and the normalization
factor should be obtained from the pair weights. However, we find that the difference
between the normalization factor obtained from the parent sample and from the pair weights
is small (a factor $\lesssim 10^{-3}$).

\subsection{Results}
\label{sec:results}
We run the fibre assignment algorithm (Section~\ref{sec:fib_assign}) 
2048 times on the BGS mock in order to
generate weight vectors for each galaxy. In each realization, a random set of
10\% of the priority 2 galaxies are promoted to priority 1, and the tile positions
are randomly dithered by an angle 3 times the patrol radius 
($3 R_\mathrm{patrol} = 4.45$ arcmin). 
We apply corrections to
the clustering measured from two volume limited samples, defined in Table~\ref{tab:samples}.
The maximum redshift of the main sample is chosen such that the sample only contains
priority 1 galaxies, while the maximum redshift is increased for the extended sample so 
that it also includes priority 2 galaxies. 
The number densities of the two samples differ slightly, due to evolution of the number
density with redshift in the mock.

\begin{table} 
\caption{Definition of the main and extended volume limited samples. Both samples use 
the magnitude range $-22<M_r - 5\log h <-21$, where the absolute magnitudes are
in the DECam $r$-band, and k-corrected to $z=0.1$.
$z_\mathrm{min}$ and $z_\mathrm{max}$ are the minimum and maximum redshifts,
$N_\textrm{gal}$ is the total number of galaxies is the sample, $f_\textrm{P1}$ 
is the fraction of priority 1 galaxies, and $\bar{n}$ is the average number density.}
\begin{tabular}{cccccc}
\hline
sample & $z_\mathrm{min}$ & $z_\mathrm{max}$ & $N_\textrm{gal}$ & $f_\textrm{P1}$ & $\bar{n}~(h^3 \Mpc^{-3})$ \\
\hline
main     & 0.09 & 0.30 &  1,532,903  & 1.00 & $1.74 \times 10^{-3}$ \\
extended & 0.09 & 0.35 &  2,655,707  & 0.94 & $1.94 \times 10^{-3}$ \\
\hline
\end{tabular}
\label{tab:samples}
\end{table}

\subsubsection{Galaxy Weights}

\begin{figure} 
\centering
\includegraphics[width=\linewidth]{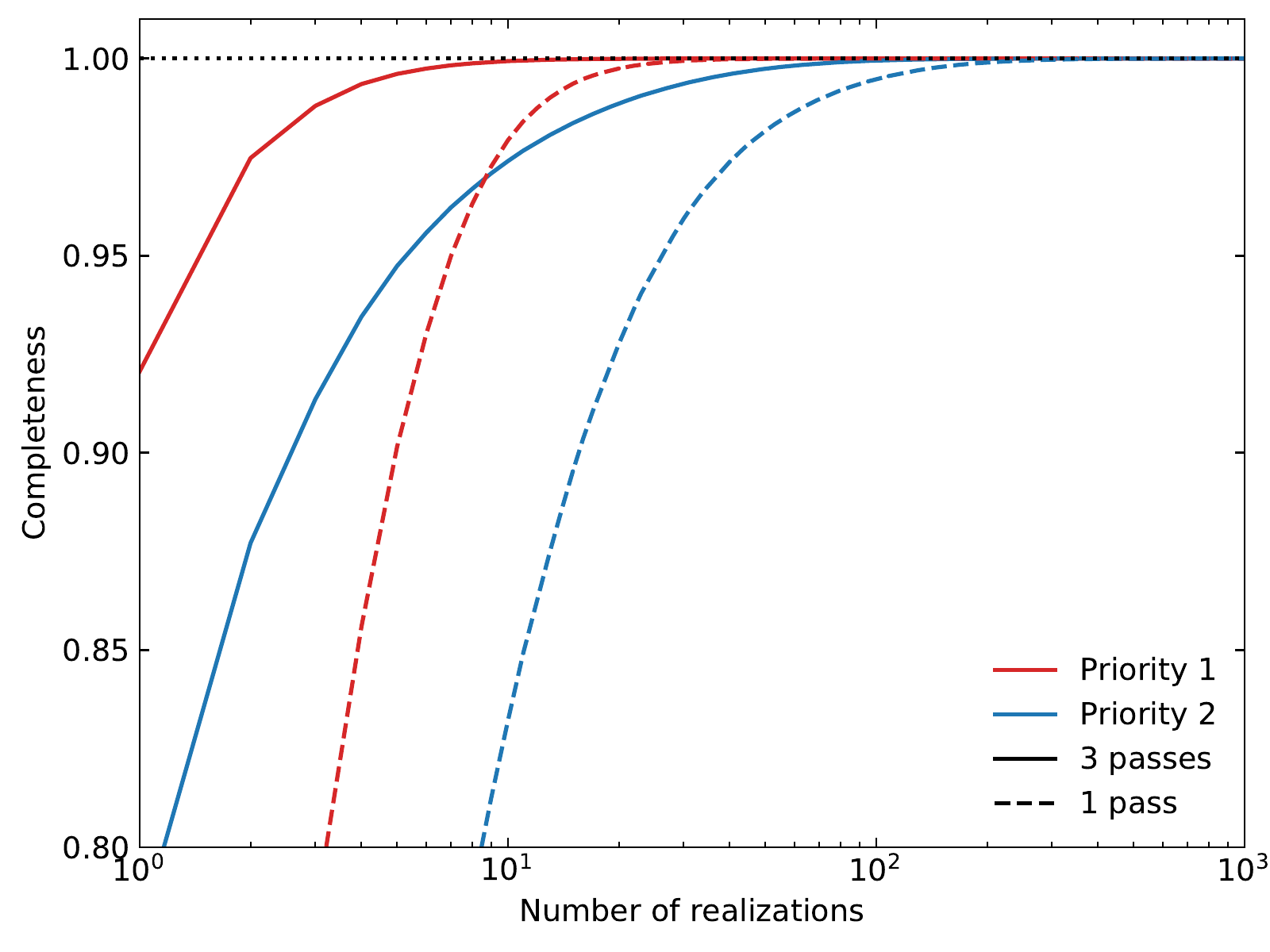}
\caption{Completeness of galaxies that are assigned a fibre at least once
after $N$ random realizations of the fibre assignment algorithm.
The full flux limited priority 1 and priority 2 samples are shown in red and blue respectively,
where solid lines are with the full 3 passes of tiles, and dashed lines a single pass.
In each realization, 
10\% of priority 2 galaxies are randomly promoted to priority 1,
and the tile centres are randomly dithered by 3 times the patrol radius.}
\label{fig:completeness_vs_realizations}
\end{figure}

The fraction of galaxies assigned a fibre at least once after $N_\mathrm{real}$ realizations of
the fibre assignment algorithm is shown in Fig.~\ref{fig:completeness_vs_realizations} for
priority 1 and 2 galaxies, with 1 and 3 passes.
To achieve a completeness of 99.99\% for
priority 1 galaxies with 3 passes, only 20 realizations are needed, while the 
same completeness for priority 2 galaxies requires around 180 realizations. 
With only a single pass of tiles, the number of realizations needed increases to
50 and 400 for priority 1 and 2 galaxies respectively. There are $\sim 10$ galaxies that are 
not assigned a fibre in any of the 2048 realizations. This
number is so small that it will have a negligible effect when applying the
pair weighting correction to clustering measurements. This number of realizations
is sufficient to estimate accurate pair probabilities for the vast majority of
galaxy pairs. However, note that the number of galaxies with zero probability, can
only be used to infer a lower bound for the number of zero probability pairs.

\begin{figure} 
\centering
\includegraphics[width=\linewidth]{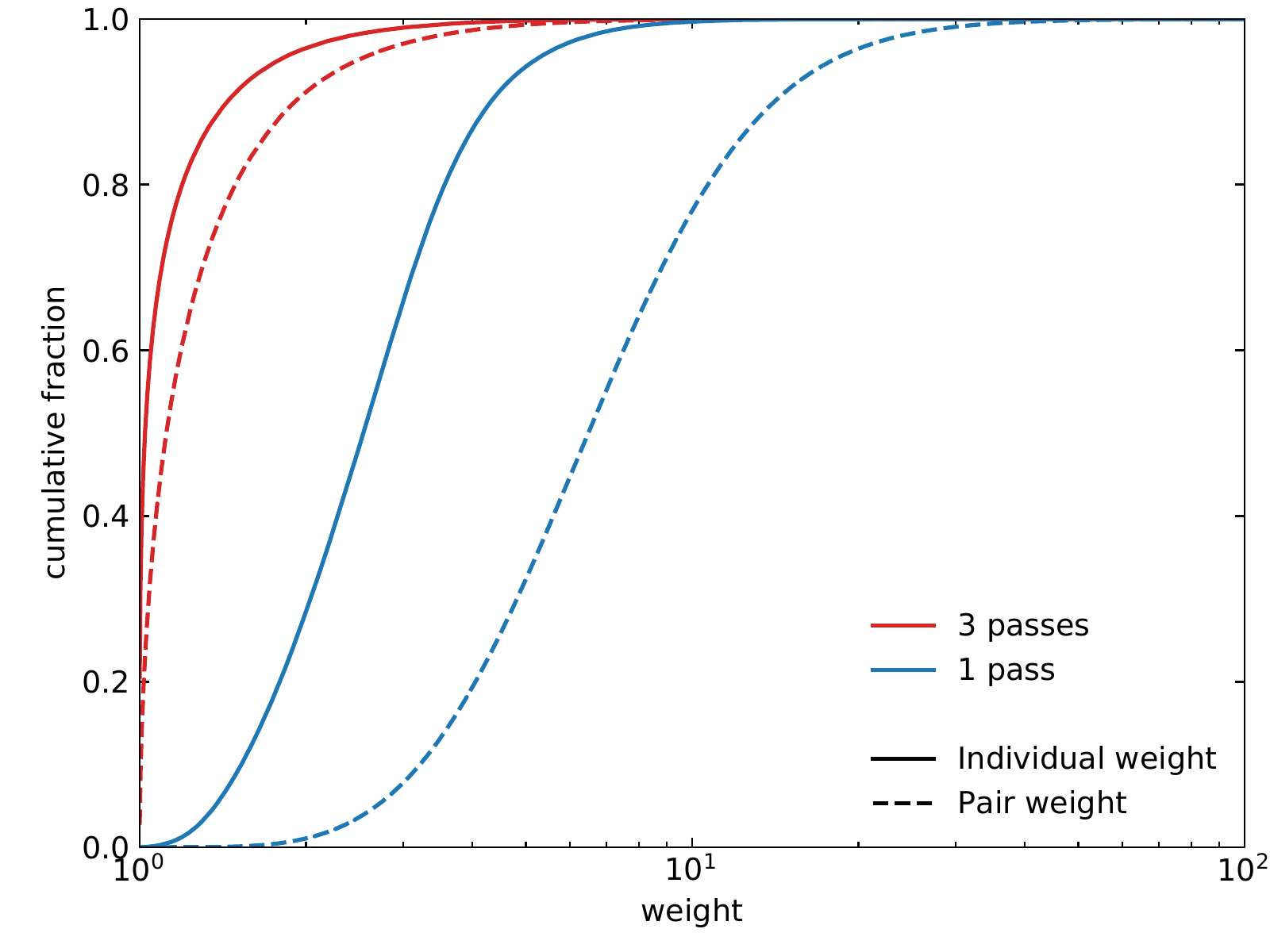}
\caption{Cumulative distribution of individual galaxy weights (solid curves) 
and pair weights (dahsed curves) of objects in the main 
volume limited sample with 1 (blue) and 3 (red) passes of tiles. For the individual weights,
the median, 90th and 99th percentiles are 1.03, 1.44 and 3.04 respectively with 3 passes,
and 2.54, 4.33 and 7.70 with a single pass. The same percentiles for the pair weights are
1.12, 1.91 and 4.39 (3 passes) and 6.50, 14.12 and 29.68 (1 pass).
After 3 passes. 16\% of objects are targeted in every realization, and have a weight exactly equal to 1,
while 2.7\% of pairs are targeted in every realization.}
\label{fig:weights}
\end{figure}

The distribution of IIP and PIP weights for the main volume limited 
sample is shown in Fig.~\ref{fig:weights}. Most of the priority 1 galaxies
are targeted in every fibre assignment realization, and so the distribution of individual weights peaks
at unity, with a tail extending to higher weights, due to objects in regions
around the edge of the survey that are only covered by a single tile and
have a low probability of being targeted. The pair weight distribution has a similar
shape, but extends to higher weights.
With only one pass, this distribution is very different,
since $\sim 90\%$ of the survey is covered by a single tile.
There are no objects targeted in every realization, and the individual weight distribution 
peaks at weight $\sim 2$, while the pair weight distribution peaks at $\sim 5$, with a 
tail extending out to very large weights.

\begin{figure}
\centering
\includegraphics[width=\linewidth]{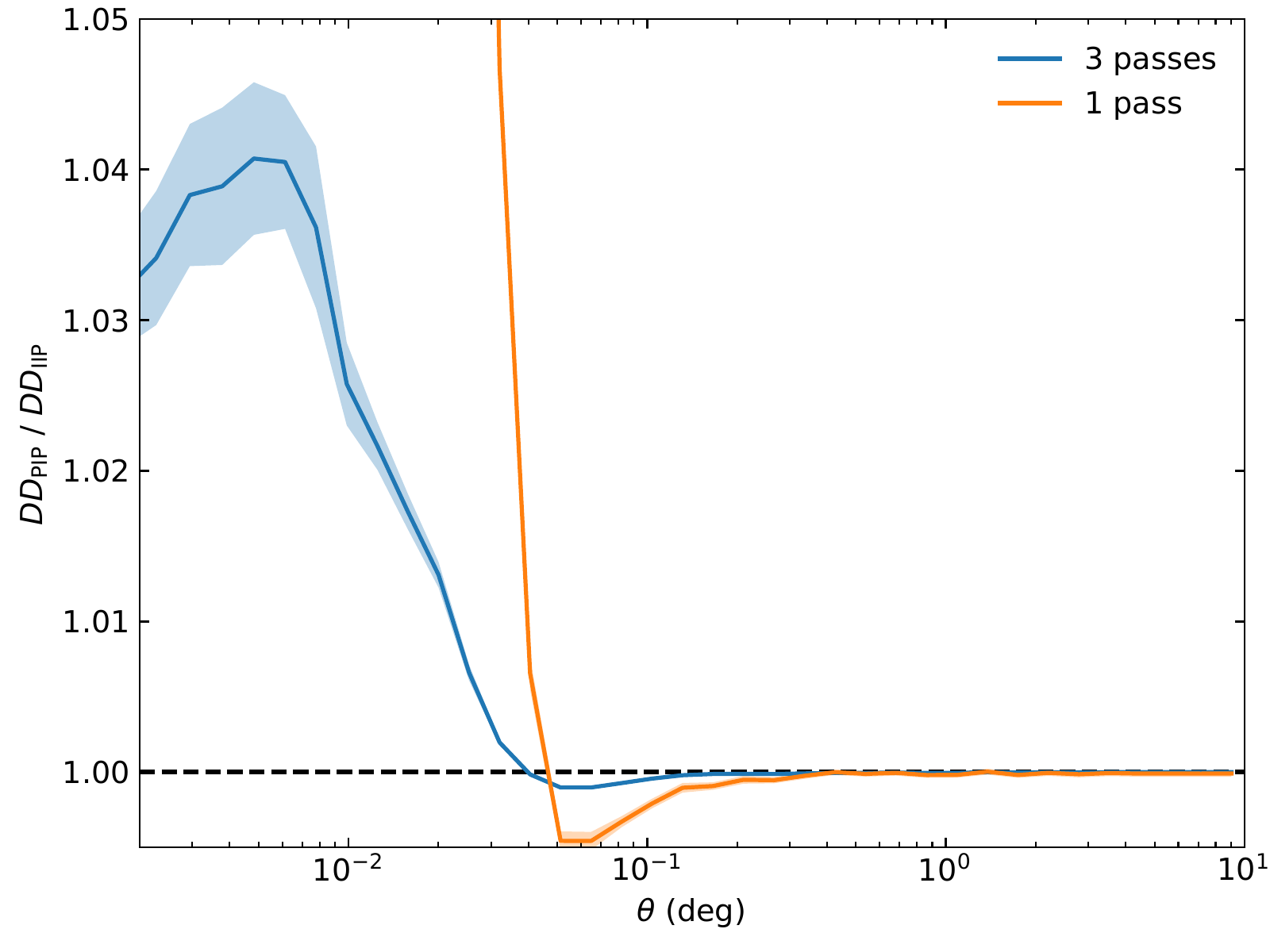}
\caption{Ratio of angular DD counts calculated with pairwise, PIP, weights to that with individual
IIP weights,
for galaxies in the main volume limited sample, after the full 3 passes of tiles (blue),
and after 90\% of 1 pass (yellow).
The solid curves are the average of 50 fibre assignment realizations, where the shaded regions
indicate the $1 \sigma$ scatter. The black horizontal dashed line indicates
a ratio of unity. The ratio on small scales after 1 pass is $\sim 1.8$.}
\label{fig:correlations}
\end{figure}

Fig.~\ref{fig:correlations} shows the ratio of the total DD counts in angular bins with PIP and
IIP weights, for the main volume limited sample, after 1 and 3 passes, illustrating 
how the correlation between pairs varies as a function of angular separation. 
On small scales, this ratio is greater than 1, indicating that the targeting probabilities are 
correlated, and $w_{ij}>w_i w_j$. At intermediate scales, there is a small negative
correlation, which asymptotes towards 1 on large scales, where $w_{ij} \sim w_i w_j$. 
However, even at 10 deg, there is a very weak correlation, and the ratio is offset
from 1 by $\sim 10^{-5}$. The size of the small scale
correlation depends on the galaxy sample and number of passes. After 3 passes, the
DD counts differ by $\sim 4\%$. After only
single pass, since most of the area has single tile coverage, correlations
are much larger, and the ratio of DD counts is $\sim 1.8$.

\subsubsection{Comparison of mitigation techniques}
\label{sec:mitigation_comparison}

\begin{figure} 
\centering
\includegraphics[width=\linewidth]{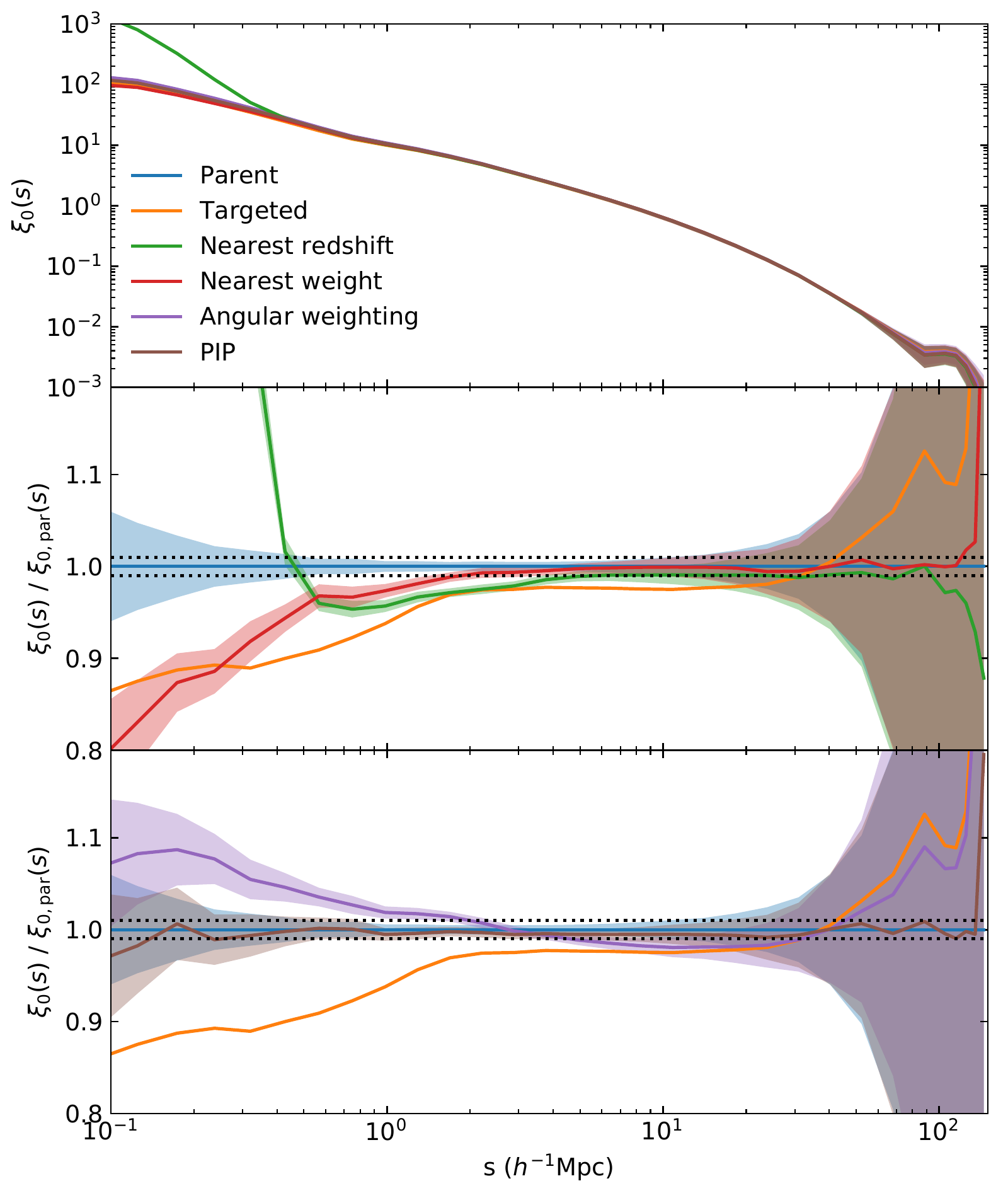}
\caption{Monopole of the redshift space galaxy correlation function of the main volume limited sample, 
with different corrections applied. The complete parent sample is shown in blue, 
targeted with no correction
in yellow, assigning missing galaxies the redshift of the nearest targeted 
galaxy on the sky in green, transferring the weight of missing galaxies to the 
nearest targeted galaxy in red, angular upweighting in purple, and PIP weighting
in brown. \red{The ratio to the complete parent sample, for different
correction methods, is split between the two lower panels for clarity.}
Shaded regions are errors estimated from 100 jackknife samples. Horizontal black dotted
lines indicate $\pm 1 \%$. For $s \gtrsim 20 \hMpc$, the scatter is almost the same for 
all methods.}
\label{fig:monopole}
\end{figure}

Fig.~\ref{fig:monopole} compares the results of applying several commonly used
correction methods
to the monopole of the redshift space correlation function of the main 
volume limited sample, after 3 passes. Each correction is
applied to a single realization of the fibre assignment algorithm, and 
errors are estimated from 100 jackknife samples (see Fig.~\ref{fig:footprint}).
The jackknife error is an estimate of the uncertainty in the clustering measurements
due to the finite survey volume.
The data is split into
100 regions of equal area, and the correlation function is calculated 
with each region omitted. The jackknife errors are taken from the square root 
of the diagonal terms of the covariance matrix. The ratio
to the complete parent sample is shown in the lower panel. The purple curve
shows the result of applying angular weighting, which by construction, reproduces
the angular correlation function of the parent sample. However, this does
not provide a satisfactory correction to the monopole. At scales of $\sim 10\hMpc$,
it differs from the parent sample by $\sim 2\%$, which is approximately twice 
the statistical error in the complete sample. At small scales, close
to $0.1 \hMpc$, it differs by almost 10\%, while the statistical error in the parent sample
is $\sim 5\%$.

Assigning missing objects the redshift of the closest targeted object on the
sky, shown by the green curve in Fig.~\ref{fig:monopole}, 
does better than angular weighting at large scales, correcting the 
monopole to a level of $\sim 1\%$. However, this correction produces
a strong artificial boost to
the clustering at small scales. Some of the untargeted galaxies will be
members of clusters, and if the nearest targeted object is also a member
of the same cluster, the redshift it is assigned will be close to the true
redshift. However, if two galaxies at different redshifts are close together 
on the sky by chance, the error in the assigned redshift could be large.
This chance projection of galaxies boosts the redshift space monopole at 
$0.1 \hMpc$ by an order of magnitude. 

Transferring the weight of missing galaxies to the nearest targeted galaxy on
the sky, which is shown by the red curve in Fig.~\ref{fig:monopole}, 
produces a correction at large scales that is within 1\%. 
The total weight of galaxy clusters is correct,
and so the large scale clustering agrees with the parent sample. However, 
since small separation pairs are missing, the clustering on small scales is
low.

The PIP correction, shown by the brown curve in Fig.~\ref{fig:monopole}, 
produces a correction within $\sim 1\%$ at all scales, even on small scales below a few
$\hMpc$ where other correction methods fail. Here, the correction is only applied
to a single fibre assignment realization, but in the next section we apply the same correction to
many realizations to check that is unbiased.

Note that only the monopole is shown in Fig.~\ref{fig:monopole}. We show in
Section~\ref{sec:multipoles} the the PIP scheme also works well for the quadrupole
and hexadecapole. The other correction methods explored in this section
fare less well for the higher order multipoles, only showing
agreement with the parent sample on scales larger than a few 10s of $\hMpc$.

\begin{figure} 
\centering
\includegraphics[width=\linewidth]{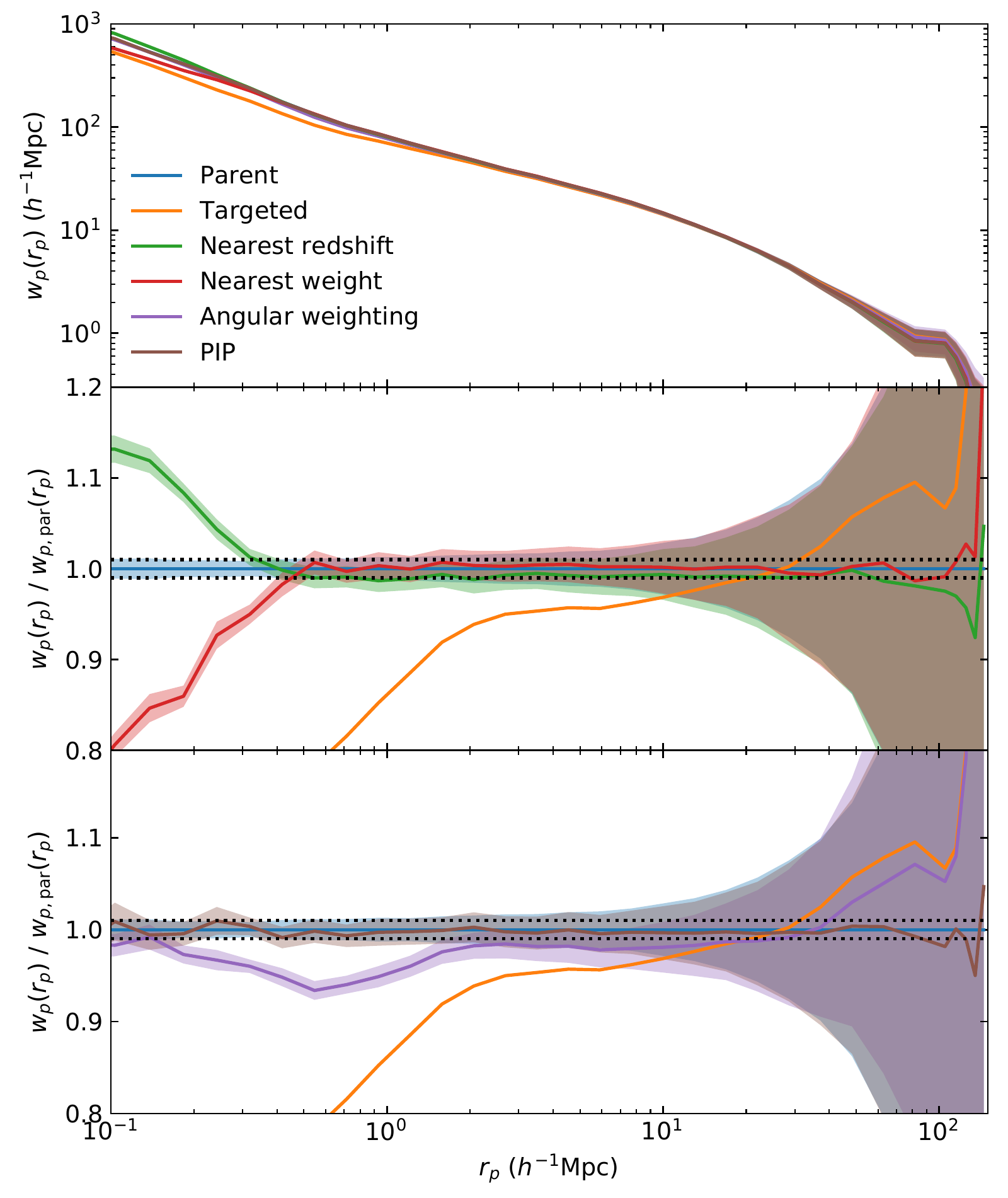}
\caption{Projected correlation function of the main volume limited sample,
with the same corrections applied as Fig.~\ref{fig:monopole}. Shaded regions
are errors estimated from 100 jackknife samples.}
\label{fig:wp_rp}
\end{figure}

The projected correlation function, 
\begin{equation}
w_p(r_p) = 2 \int_0^{\pi_\mathrm{max}} \xi(r_p, \pi) d\pi, 
\end{equation}
is shown in Fig.~\ref{fig:wp_rp}, with the same corrections applied, and using 
$\pi_\mathrm{max} = 120 \hMpc$. The two nearest redshift corrections are able to
correct the projected correlation function to within 1\% down to a scale of 
$\sim 0.5 \hMpc$. Since the projected correlation function integrates along the
line of sight, it reduces the impact of galaxies which are assigned the wrong
redshift. Again, the PIP weighting produces a correction to within $\sim 1\%$
on all scales.

\red{Another correction method we have not considered here is to
down-weight objects in the random catalogue by the probability that
a galaxy in that location could be targeted. However, this correction will only be unbiased
if the completeness of galaxies is uncorrelated with density, which is not
true in the BGS. The randoms would be down-weighted in high density regions,
giving these regions less weight, and producing a biased estimate of the correlation 
function. \cite{Pinol2017} measure the power spectrum using weighted random catalogues, and
show that it is unable to produce an adequate correction to the power spectrum, 
without removing low $\mu$ bins.}

\subsubsection{Angular clustering with PIP weights}

\begin{figure*} 
\centering
\includegraphics[width=0.65\linewidth]{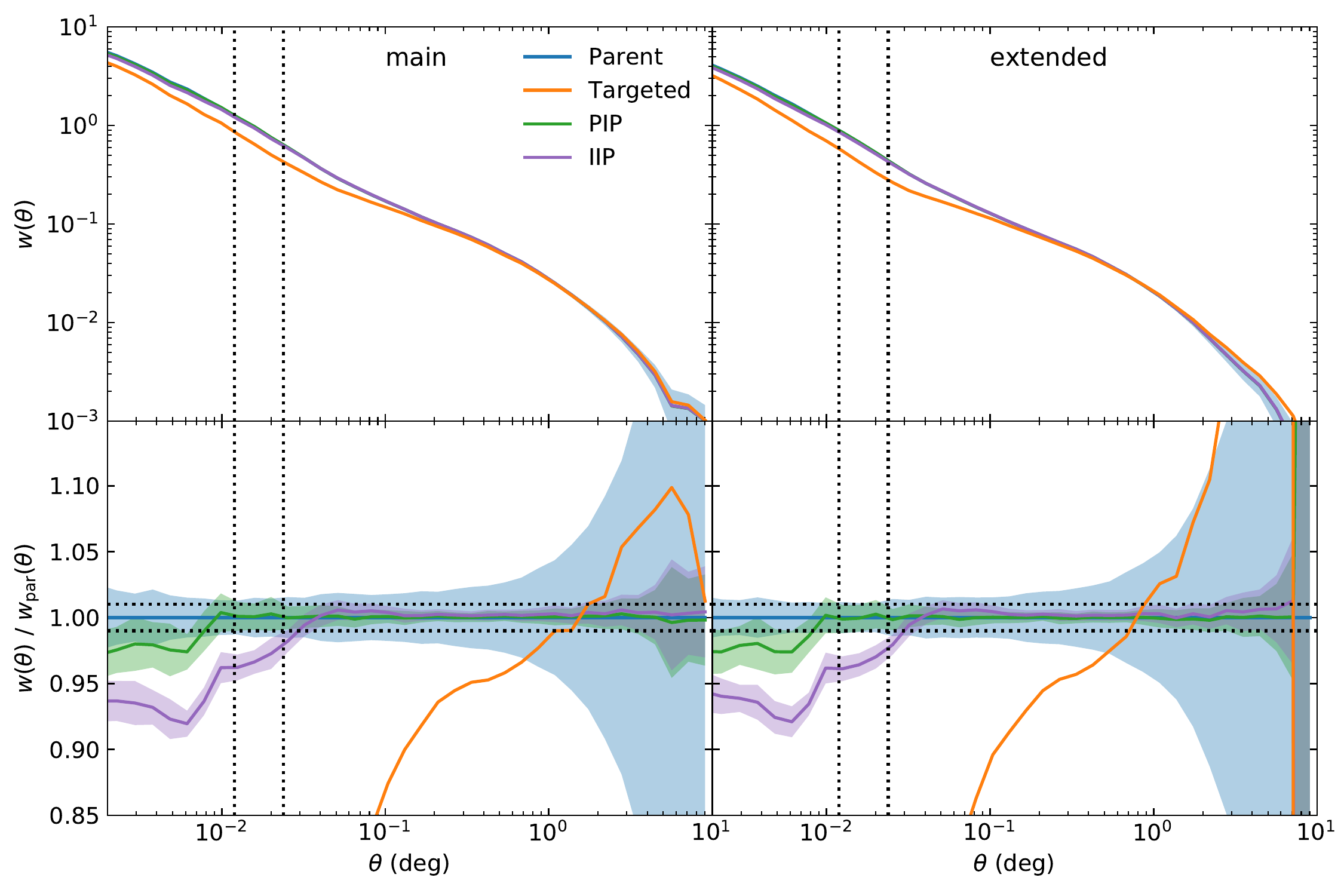}
\caption{Angular correlation function for the main volume limited sample that only contains
priority 1 galaxies (left), and the extended volume limited sample that also contains 
priority 2 galaxies (right), after the full 3 passes of tiles. 
The bottom panels show the ratio to the complete parent sample.
The parent sample is shown in blue, where the shaded region indicates the error from
100 jackknife samples. The yellow curve illustrates the angular correlation function from
one realization of fibre assignment, with no correction. Green and purple curves are the results
of applying pair weighting and individual galaxy weighting, respectively, averaged over
50 realizations. The shaded regions indicate the scatter between these 50 realizations.
Vertical dotted lines indicate the angular scale of $R_\mathrm{patrol}$ and $0.5 R_\mathrm{patrol}$
and the horizontal lines indicate $\pm 1\%$.}
\label{fig:wtheta_3pass}
\end{figure*}

We now apply the PIP weighting to the angular correlation function. 
By construction, the angular correlation function of the parent sample is recovered
exactly when the pair weighting and angular correction of Eq.~\ref{eq:dd_pip} are both applied.
However, it is interesting to see how well the PIP weighting on its own can recover the angular
correlation function for a volume limited sample, where in the real survey, the complete
parent sample would not be known.
To check that the correction is unbiased, we average
the result of applying the correction to 50 fibre assignment realizations (which are a subset
of the 2048 realizations used to estimate the pair weights). 
The result, after 3 passes, is shown in Fig~\ref{fig:wtheta_3pass}.
The left panels show the angular
correlation function of the main volume limited sample, with the ratio to the complete 
parent sample in the bottom panel. 
The parent sample is shown in blue, where the shaded region is the statistical
error, estimated from 100 jackknife samples.
The yellow curve shows the correlation function of galaxies assigned 
fibres in a single realization of fibre assignment, illustrating the size of the 
correction that needs to be made. 
The green curve illustrates the result of applying only the pair weighting, without
the angular upweighting term, and is the mean of 50 realizations of fibre assignment.
The shaded 
region indicates the $1\sigma$ scatter between these realizations. 
This is the additional error due to measuring the clustering from a subset of the
objects in the parent sample, and we aim for this to be small compared to the statistical error
in the parent sample.
On large scales, the
pair weighting does an excellent job of correcting the angular clustering. The mean
is unbiased, and the scatter is within 1\% for angular scales between $\sim 0.03$ deg 
and 1 deg. This is much smaller than the statistical error in the parent sample,
which is of the order of a few percent, increasing on larger scales. 
However, on small scales, less than $0.5 R_\mathrm{patrol}$, there is a small
bias of a few percent. This bias is due to pairs of galaxies around the edge of the 
survey, in regions covered by only a single tile. Pairs of galaxies with a very small
angular separation in these regions can never be targeted due to fibre collisions, even
when the tiles are dithered. Since these pairs have a zero probability of being targeted,
this results in a bias, which is corrected for by the angular upweighting term.
It is not guaranteed that that this angular correction will be accurate since, for example,
missing pairs could occur preferentially in triplets, and therefore be statistically 
distinct from targeted pairs of the same separation. However, we find that this is not
the case, and the missed pairs fall in the regions of single tile coverage.
Alternatively, the edge of the survey could be trimmed, removing the regions covered by a single tile,
which is only a small percentage of the footprint ($\sim 3\%$, see Table~\ref{tab:overlaps}).
Another alternative strategy is discussed in Section~\ref{sec:discussion}.

For comparison, the purple curve shows the result of applying individual galaxy weights
to the same set of realizations.
At small scales, applying individual weights results in a larger bias than pair weights,
and this bias extends to larger angular scales. This is because individual galaxy weights
do not take into account any correlation between galaxy pairs. For example, if it is difficult to
target both galaxies in a pair due to fibre collisions, but relatively easy to target
one or the other individually, calculating the pair probability from individual probabilities
is biased since $p_i p_j > p_{ij}$.
On large scales, if there are no correlations between pairs, $p_i p_j = p_{ij}$, and using individual
weights should produce the same result as pair weights. However, in Fig~\ref{fig:wtheta_3pass}, 
there is still a small difference between the green and purple curves on large scales. 
Even at scales of $\sim 10$ deg, there is still some 
correlation between galaxy pairs, although this is very small, with a fractional difference
in the $DD$ counts of $\Delta DD / DD \sim 10^{-5}$. The fractional error in $\xi$ is given by
\begin{equation}
\frac{\Delta \xi}{\xi} \approx \frac{\Delta DD}{DD} \frac{(1+\xi)}{\xi}.
\end{equation}
On large scales, $\xi \sim 10^{-3}$, which results in a fractional difference of $\Delta \xi/\xi \sim 1\%$,
which is a small, but noticeable difference in the correlation function.

The right hand panels of Fig.~\ref{fig:wtheta_3pass} shows the result of applying the same
corrections to the extended volume limited sample, which also contains priority 2 galaxies.
By giving the priority 2 galaxies a small probability of being promoted to priority 1,
this gives every pair of priority 2 galaxies a non-zero probability of being targeted,
and therefore applying the pair weighting correction produces an unbiased result on large
scales. There is still a small bias on small scales for the same reason as in the main sample.

\begin{figure*} 
\centering
\includegraphics[width=0.65\linewidth]{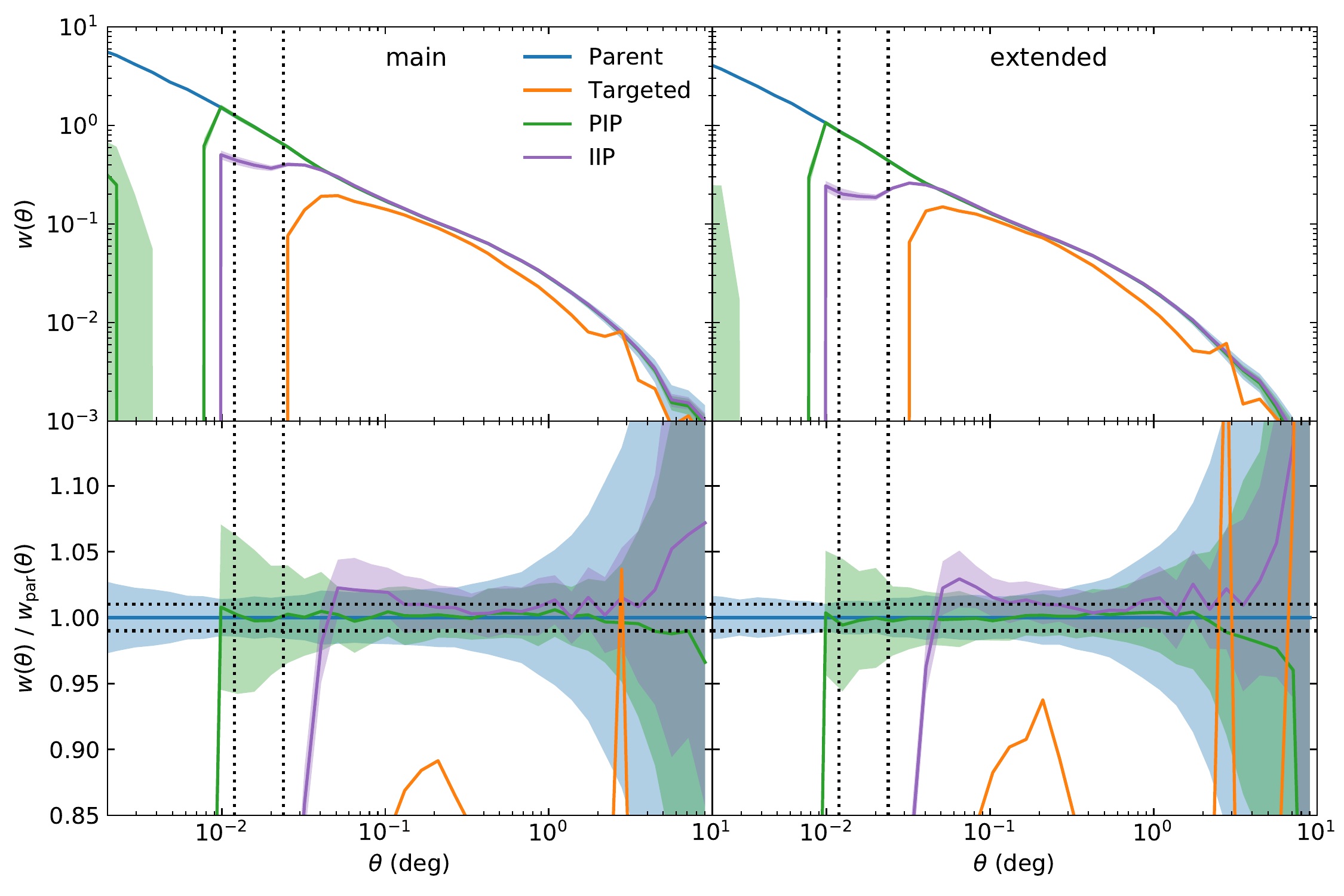}
\caption{As Fig.~\ref{fig:wtheta_3pass} but after only 1 pass of tiles, and with 10\% of the tiles missing.
This illustrates the data that might have been obtained after one third of the complete survey,
with a survey strategy that prioritised area over completeness.}
\label{fig:wtheta_1pass}
\end{figure*}

Fig.~\ref{fig:wtheta_1pass} shows the angular correlation function after only a single 
pass of tiles, with a random 10\% of the tiles missing, for the same volume limited samples. 
With only 1 pass of tiles the catalogue of fibre assigned galaxies is much less complete,
and a larger correction is required.

Since most of the footprint is covered by a single tile ($\sim 90\%$, see 
Table~\ref{tab:overlaps}), the bias on scales
less than $0.5 R_\mathrm{patrol}$ is much larger than after 3 passes. Since there
are overlaps between neighbouring tiles, the pair counts on these scales are low, but
not zero. Pair weighting must be combined with angular upweighting in order to correct
the clustering on these scales. 
%\textbf{JH:Indeed, we can use truncated multipoles instead of the normal ones.}

On larger scales, pair weights on their own are able to produce an unbiased correction,
although the scatter between realizations is larger than with 3 passes, but 
on scales above
1 degree this scatter is approximately half of the statistical error of the parent sample.

\subsubsection{Correlation function multipoles with PIP weights}
\label{sec:multipoles}

\begin{figure*} 
\centering
\includegraphics[width=\linewidth]{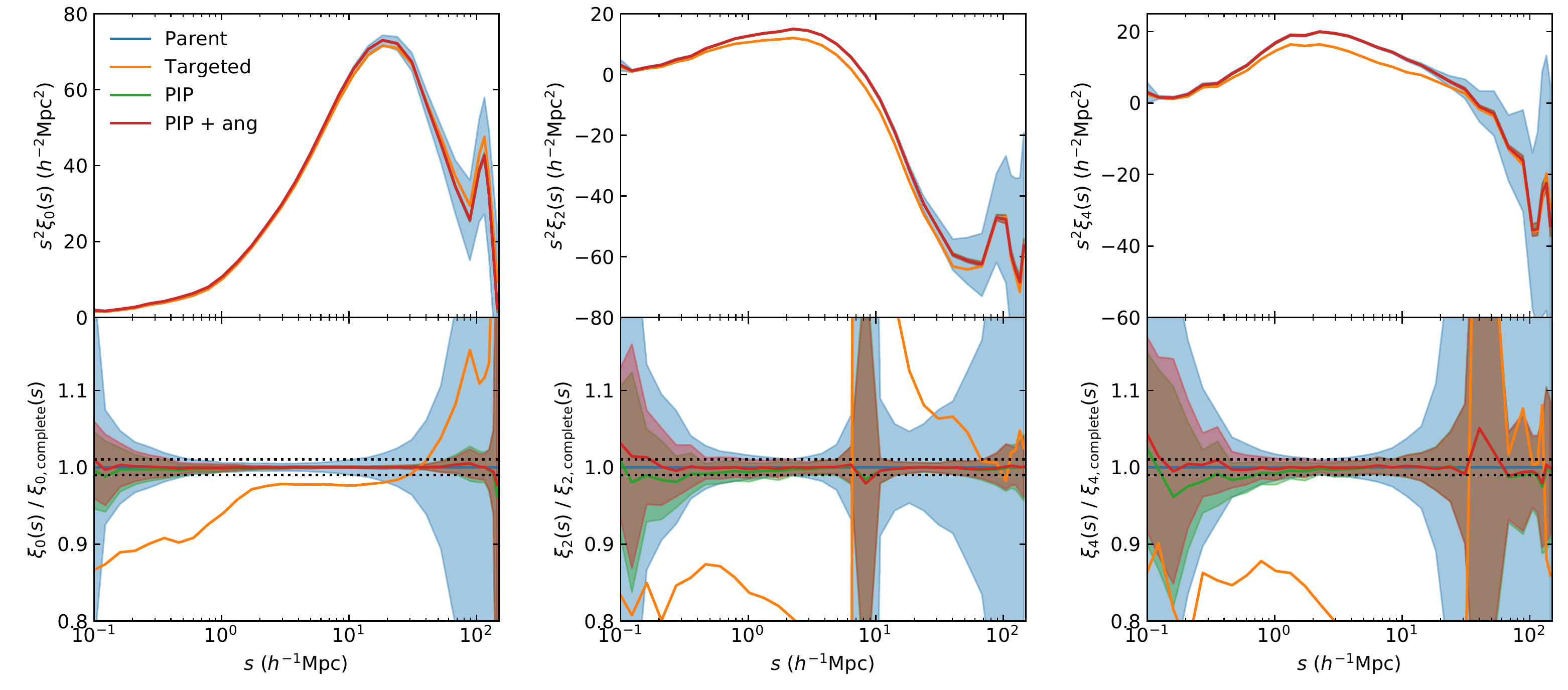}
\caption{Monopole, $\xi_0(s)$, quadrupole, $\xi_2(s)$, and hexadecapole, $\xi_4(s)$, 
of the redshift space galaxy correlation 
function for the main volume limited sample. The ratio to the complete parent sample is 
shown in the bottom panel.
The parent sample is indicated by the blue curve, where the shaded blue region
is the error from 100 jackknife samples. The green curve is the average of 50
realizations, corrected with only PIP weighting. The red curve is corrected using
both PIP and angular weighting. Shaded green and red regions indicate $1 \sigma$,
estimated from the scatter between the 50 realizations.}
\label{fig:multipole_main_3pass}
\end{figure*}

The Legendre multipoles of the redshift space correlation function for the main sample after 
3 passes are shown in Fig.~\ref{fig:multipole_main_3pass}. 
At large scales, the PIP weighting on its own is unbiased and does a good job of correcting
the measured clustering. Between 1 and 10 $\hMpc$, the scatter between realizations 
in the monopole is well within
1\%, and even for the hexadecapole the scatter is around 1\%. Note that the scatter 
in the quadrupole and hexadecapole appears to be large at $\sim 1 \hMpc$ and $\sim 5 \hMpc$
respectively, but this is just because the curves in the upper panels go through zero.

On small scales, similarly to what was seen in the angular correlation function,
applying the PIP weighting on its own produces a biased result, due to pairs that
cannot be targeted in regions covered by a single tile. Most of this
area covered by a single tile is located around the edge of the footprint.
We again find that including the angular weighting term corrects for this small bias.

\begin{figure*} 
\centering
\includegraphics[width=\linewidth]{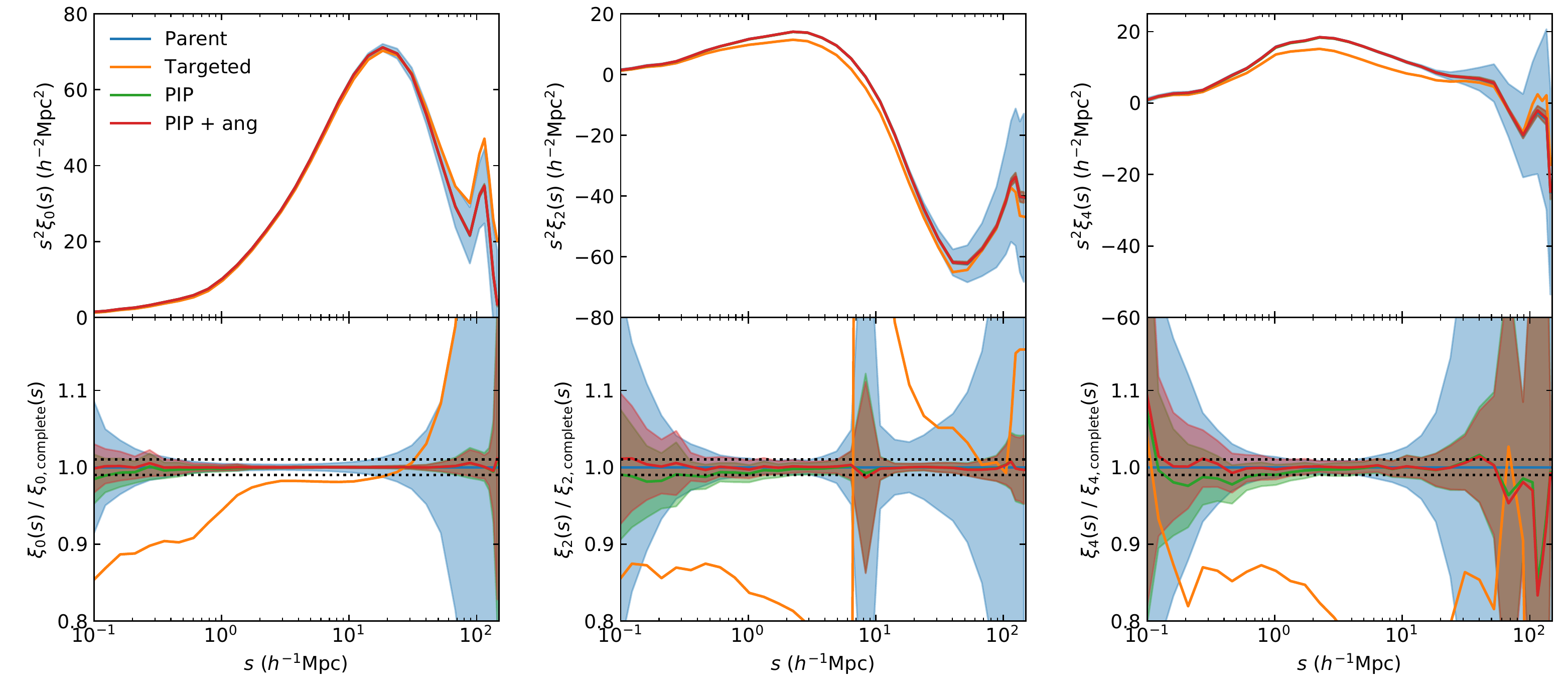}
\caption{As Fig.~\ref{fig:multipole_main_3pass}, but for the extended volume limited
sample, which contains both priority 1 and 2 galaxies}
\label{fig:multipole_extended_3pass}
\end{figure*}

Fig.~\ref{fig:multipole_extended_3pass} is the same, but for the extended sample.
The results look similar to that of the main sample, showing that including priority 2
galaxies does not produce any biases.

\begin{figure*} 
\centering
\includegraphics[width=\linewidth]{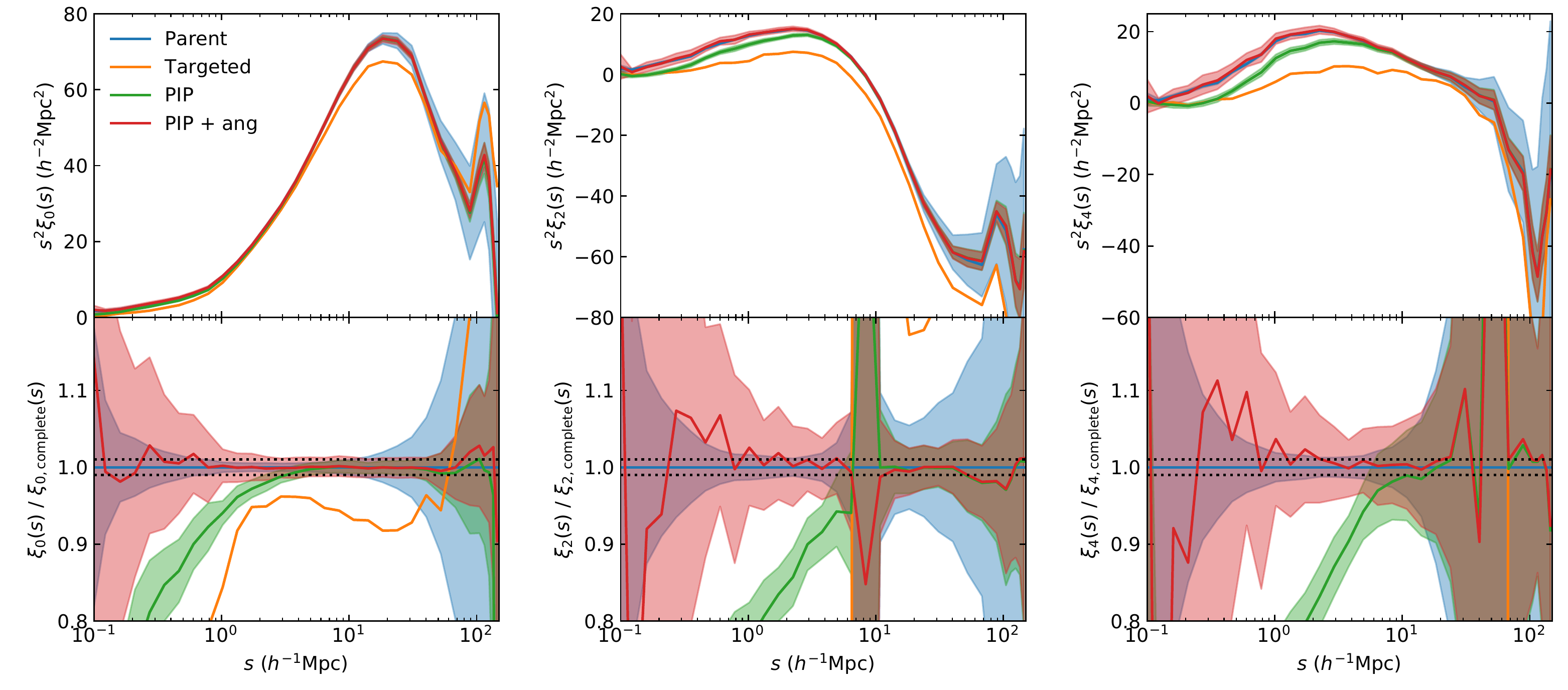}
\caption{As Fig.~\ref{fig:multipole_main_3pass}, but for the case of only a single pass
of tiles.}
\label{fig:multipole_main_1pass}
\end{figure*}

\begin{figure*} 
\centering
\includegraphics[width=\linewidth]{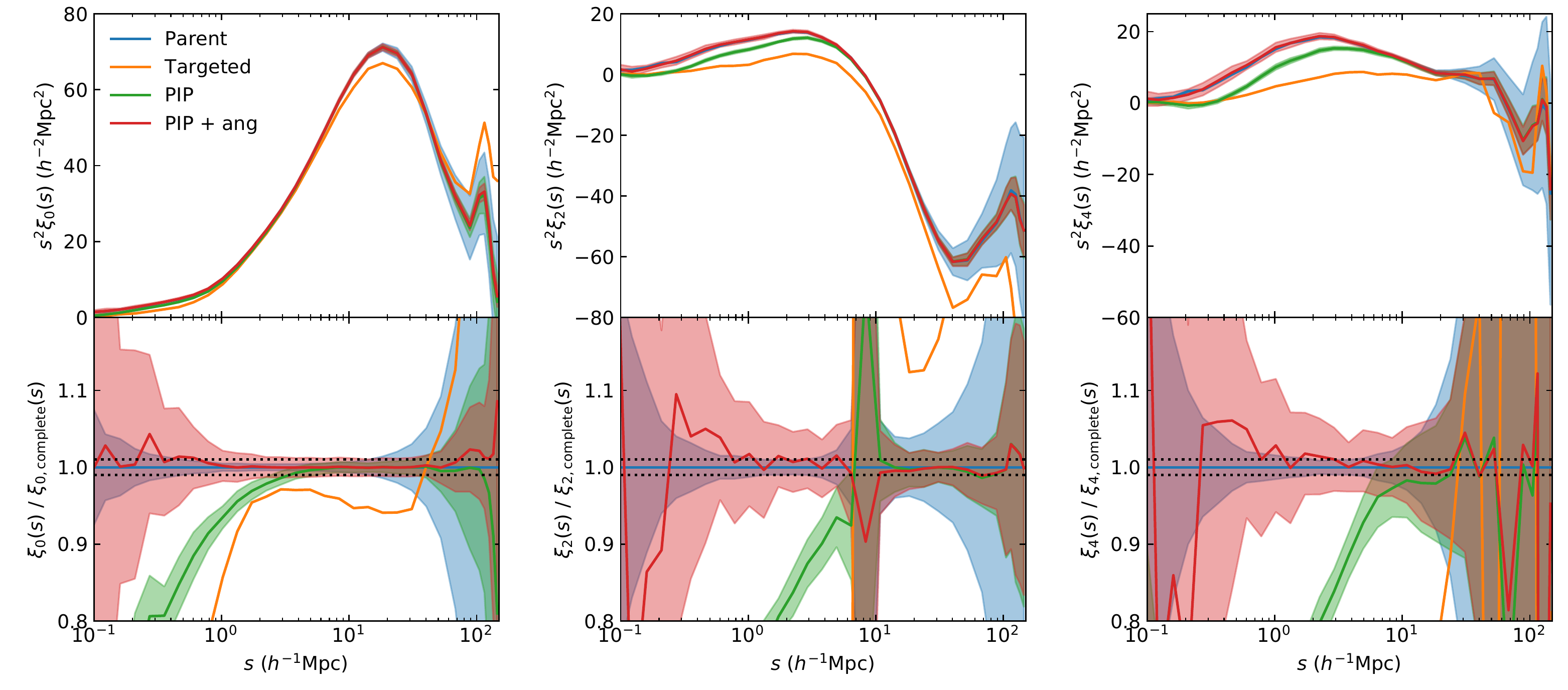}
\caption{As Fig.~\ref{fig:multipole_main_3pass}, but for the extended volume limited
sample, after only a single pass of tiles.}
\label{fig:multipole_extended_1pass}
\end{figure*}

Figs.~\ref{fig:multipole_main_1pass}~and~\ref{fig:multipole_extended_1pass} show
the results of applying the same corrections to the same volume limited samples, 
but with only 90\% of 1 pass of tiles. Since the survey is much more incomplete, the 
correction that must be applied is larger. On large scales, applying the PIP
weights on their own produces an unbiased correction, but with larger scatter
between fibre assignment realizations compared to the 3 pass case. On small scales, the bias
is much larger for PIP alone, but combining with angular weighting is able 
to correct this large small scale bias to within the errors.

After the full 3 passes of tiles, the scatter between realizations is much 
smaller than the statistical error in the parent sample on all scales. 
With only a single pass, this scatter is much larger, and on small
scales becomes larger than the statistical error.
The scatter is large after 1 pass because the sample is highly incomplete
(e.g. for the main volume limited sample, $\sim 38\%$ of objects are assigned
a fibre in each realization), and most objects have a large weight
(the median weight is 2.54, see Fig.~\ref{fig:weights}). After 3 passes,
the scatter is much smaller, since the completeness of the main sample is much higher 
($\sim 82\%$), and most objects have a weight close to unity. 
90\% of the 1 pass survey area 
is covered by a single tile, and the completeness of close pairs is very low, due to
fibre collisions. Each pair will also have a very large weight, which results in the very large 
scatter on small scales. The completeness of pairs on small scales is much higher with multiple 
passes, and therefore the scatter is much smaller.

While the average of many fibre assignment realizations 
is unbiased, the real survey is only a single realization, and after 1 pass
it is likely that there will be a large scatter between the corrected 
clustering measurements and the true clustering at small scales.
Multiple passes are therefore necessary in order to obtain precise
clustering measurements on these scales.
On large scales, the scatter is smaller than the statistical error after 1 pass,
so it will be possible to make precise BAO and large scale RSD measurements. 
However, the uncertainty in these measurements will be greatly reduced after the
subsequent passes. Multiple passes will also reduce the incompleteness due to
redshift measurement failures, as it will give these galaxies another chance
to be targeted. 
To make precise small scale RSD measurements, a single pass is not sufficient.

The shot noise in these galaxy clustering measurements could potentially be reduced 
by capping the pair weights at some maximum value. Strictly speaking, the PIP weighting
would no longer be unbiased, but this bias can be reduced by the angular weighting term,
using these capped weights. We find that for the main sample after 1 pass, capping the
weights at a maximum value of 100 (0.01\% of pairs) has a negligible affect on the monopole,
but reduces the scatter in the quadrupole and hexadecapole at scales of $\sim 1 \hMpc$
by a few percent. Capping the weight at 25 ($\sim 2\%$ of pairs) introduces systematics,
which are not corrected for completely by the angular weighting. On large scales, there is
a negligible change in the scatter, and the small bias that is introduced is within the errors.
On small scales, this bias is larger, but is still within the large errors.

\subsection{Discussion}
\label{sec:discussion}

We have shown in the previous section that the PIP weighting scheme, in combination
with angular upweighting, is able to produce an unbiased correction to clustering 
measurements in the BGS, even for a highly incomplete survey. 

One simplifying assumption we have made is that the galaxies in the parent sample 
are known.
The angular weighting term from Eq.~\ref{eq:dd_pip} includes $DD^{(p)}$,
the angular data-data pair counts of the complete parent sample (and similarly for 
the DR counts, $DR^{(p)}$). For a flux-limited sample, the parent sample is known,
but this is not true in the case of a volume limited sample, since every redshift
would need to be measured to determine an absolute magnitude, and hence which galaxies belong 
in the sample.
When applying the angular weighting, we have used the true parent sample, which in the 
real survey would not be known. %We leave this for future work to investigate
%how well the correction works when the parent sample is not known.

In order to calculate pair weights, we dither
the catalogue by a small angle in each realization of fibre assignment. For galaxies close to the edge of
the survey, in half of the realizations they will fall outside the footprint, which
results in these galaxies having larger weights than galaxies in the centre. 
In the actual survey, the dither is zero, which is a special case where no objects
fall off the edge, and is not strictly represented in the ensemble of realizations.
However, we find no measurable bias as only a very small
fraction of objects are affected.

An issue that affects the real survey that we have not considered is stellar contamination.
A small fraction of objects in the catalogue of potential targets are stars that have been
misclassified as galaxies. If a fibre is placed on one of these objects, and a spectrum measured,
it can be determined that it is a star and not a galaxy.
Since the PIP weighting scheme
can produce an unbiased correction to clustering measurements of any sub-sample of galaxies,
the misclassified stars can simply be removed when estimating the correlation function. As long as the 
stars are included when running the fibre assignment algorithm many times to estimate the PIP weights,
this will produce unbiased clustering measurements.

An alternative way to dither the catalogue would be to place the survey tiling randomly
on the full sky, with a random orientation. This has the advantage that the undithered 
catalogue is not a special case, and could be drawn from these random tile positions.
Also, every part of the sky has a non-zero probability of being in an area of the survey
covered by multiple tile overlaps, giving every pair, even at very small separations, a
non-zero probability of being targeted. This means that $w_{ij}$ pair weights without
angular weighting can produce an unbiased correction,
so the correction can be applied without knowledge of the complete parent sample. 
However, in many of these fibre assignment realizations, the tiling
would cover large areas of the sky which are outside the BGS footprint. 
Despite this, we expect that the total number of realizations needed to accurately
estimate pair weights will be smaller, since the tail of pairs with extremely high weights
are much more likely to be targeted in the realizations where they are covered by
multiple tile overlaps.
%This means that
% a greater number of realizations would be required in order to estimate the pair weights.
% Since the BGS covers around one third of the sky, 3 times the number of realizations would 
% be needed to estimate the pair weights for pairs with a small separation. For pairs with
% very large separations, many more realization would be needed.

A similar method to this is used in \citet{Mohammad2018}, where in order to estimate
pair weights for galaxies in the VIPERS survey, the parent catalogue
is rotated by angles of either 0, 90, 180 and 270 degrees, and the spectroscopic mask is moved 
spatially. The PIP weighting scheme is shown to work well, and this is the only published
example of applying the PIP weights to a real dataset.

With large dithers across the full sky, it is also necessary to modify the definition of 
pair weights to take into account that galaxies will fall outside the survey tiling
in many of these realizations of fibre assignment. Consider a perfect survey in which if two galaxies fall
within the survey tiling, it is always possible to target the pair, so all pairs should
have the same weight. If the pair have a very small angular separation, then in 1/3 of
realizations they will fall within the tiling and be able to be targeted,
so they would have a pair weight of 3, using Eq.~\ref{eq:pair_weights}. 
However, if a pair has a very large separation, it can 
be unlikely that both fall within the tiling at the same time in a random realization,
so the pair probability is low and therefore the weight will be much larger than 3.
Eq.~\ref{eq:pair_weights} incorrectly gives pairs of different 
separations different weights.
Instead, the pair weight can be redefined as
\begin{equation}
w_{ij} = \frac{\vec{c}_i \cdot \vec{c}_j}{\vec{w}_i \cdot \vec{w}_j},
\end{equation}
where $\vec{c}_i$ is a bitwise coverage vector that contains a $1$ if it is possible 
to place a fibre on galaxy $i$ (i.e.the galaxy lies within the patrol region of a fibre though 
it may happen not to be targeted) in that realization, and $0$ otherwise.\footnote{The ability to
use bitwise coverage vectors is implemented in the correlation function code \textsc{twopcf} \citep{Stothert2018}.} 
Applying this
definition in the above example results in all pairs having a weight of $1$, as
expected.

We have only shown the results of applying the correction to volume limited
samples with a number density $\sim 2\times 10^{-3} h^3 \Mpc^{-3}$. We have
also applied the correction to volume limited samples of different number 
densities, and samples defined by a colour cut, and we find that applying the PIP
correction with angular weighting will produce an unbiased correction.

\section{Conclusions}
\label{sec:conclusions}

The DESI BGS will be a highly complete, flux limited spectroscopic survey
of low redshift galaxies, an order of magnitude larger than existing galaxy catalogues,
with the primary science aims of BAO and RSD analysis. 
Fibres in the focal plane of the telescope are controlled by robotic fibre positioners,
each of which can place a fibre on any galaxy within a small patrol region, leading to  
incompleteness in the catalogue due to fibre collisions, and the fixed density of fibres 
over large regions in each tile. This leaves a non-trivial impact on clustering measurements,
and it is essential that these biases can be corrected.

We have quantified the targeting completeness in the BGS by applying the DESI fibre assignment
algorithm to a BGS mock catalogue. To ensure each galaxy has a non-zero probability
of being targeted, and to maximize the number of pairs that can be targeted,
we randomly promote 10\% of faint priority galaxies to the same priority as 
the bright priority 1 galaxies, and dither the tile positions by a small
angle of 3 times the fibre patrol radius.

The main determinant of completeness in the BGS is the surface density of
galaxies. Completeness is high in low surface density regions,
(e.g. over 95\% for priority 1 galaxies after 3 passes), but drops
significantly in the most overdense regions. Close to the centre
of the very most massive haloes ($\sim 10^{15} \hMsun$), the completeness can 
be as low as 10\% or less.

We applied several correlation function correction methods to volume limited
samples from the BGS mock catalogue, where the incompleteness is due to fibre
assignment only. This is done for a highly complete survey
with 3 passes of tiles, and a highly incomplete survey, with 1 pass
and 10\% of the tiles missing. Using standard angular upweighting, or
assigning missing galaxies the redshift of the nearest targeted galaxy
provide an unsatisfactory correction to the correlation function monopole
on small scales below a few Mpc (and a few 10s of Mpc for the
higher order multipoles).

After 3 passes of tiles, the method of \citet{Bianchi2017}, which combines
galaxy pair weights with an angular weighting,
is able to produce an unbiased correction to the angular and redshift 
space correlation functions, where the scatter between fibre assignment realizations is 
much smaller than the statistical error in the complete parent sample. 
The angular weighting term is required to correct a small bias on small
scales caused by untargetable pairs around the edge of the survey footprint.
After 1 pass, the correction is again unbiased, but the scatter between
realizations is much larger, and on small scales the method relies heavily 
on angular weighting. More than 1 pass will be needed to make precise
RSD measurements on small scales.

We propose an alternative method to dither the tiles, where the entire
survey tiling is positioned randomly on the full sky, and the pair weight
definition takes into account realizations in which objects cannot be targeted.
This has the advantage that pair weighting on its own can produce an unbiased
correction without relying on angular weighting. %, but has the disadvantage that
%many more realizations are required in order to obtain accurate pair weights.

\section*{Acknowledgements}

This work was supported by the Science and Technology Facilities Council (ST/M503472/1).
AS, JH, SC, LS, PN and CB acknowledge the support of the Science and Technology Facilities 
Council (ST/L00075X/1). JH is supported by the European Research Council (ERC-StG-716532-PUNCA) and
the Durham co-fund Junior Research Fellowship.
LS acknowledges the support of the Science and Technology Facilities Council (ST/J501013/1).
PN acknowledges the support of the Royal Society through the
award of a University Research Fellowship.
JEFR acknowledges support from COLCIENCIAS Contract No. 287-2016, Project 1204-712-50459.

This work used the DiRAC Data Centric system at Durham University, operated 
by the Institute for Computational Cosmology on behalf of the STFC DiRAC HPC 
Facility (www.dirac.ac.uk). This equipment was funded by BIS National 
E-infrastructure capital grant ST/K00042X/1, STFC capital grants ST/H008519/1 
and ST/K00087X/1, STFC DiRAC Operations grant ST/K003267/1 and Durham University. 
DiRAC is part of the National E-Infrastructure. 

%%%%%%%%%%%%%%%%%%%%%%%%%%%%%%%%%%%%%%%%%%%%%%%%%%



\bibliographystyle{mnras}
\bibliography{ref} % if your bibtex file is called example.bib

% Don't change these lines
\bsp	% typesetting comment
\label{lastpage}
\end{document}